\documentclass[12pt]{article}
\usepackage{jheppub}

\pdfoutput=1

\usepackage{amsmath,bbm,array,amsfonts,graphicx,wrapfig,lscape,float,mathtools,multirow,longtable}
\usepackage[dvipsnames]{xcolor}
\usepackage{array}

\newcommand{\be}{\begin{equation}}
\newcommand{\ee}{\end{equation}}
\newcommand{\beq}{\begin{equation}}
\newcommand{\beql}[1]{\begin{equation}\label{#1}}
\newcommand{\eeq}{\end{equation}}
\newcommand{\ba}{\begin{array}}
\newcommand{\ea}{\end{array}}
\newcommand{\bea}{\begin{eqnarray}}
\newcommand{\beal}[1]{\begin{eqnarray}\label{#1}}
\newcommand{\eea}{\end{eqnarray}}
\newcommand{\ben}{\begin{enumerate}}
\newcommand{\een}{\end{enumerate}}
\newcommand{\bean}{\begin{eqnarray*}}
\newcommand{\eean}{\end{eqnarray*}}
\newcommand{\eref}[1]{(\ref{#1})}
\newcommand{\sref}[1]{\S\ref{#1}}

\newcommand{\fref}[1]{Figure \ref{#1}}
\newcommand{\btab}[1]{\begin{tabular}{#1}}
\newcommand{\etab}{\end{tabular}}

\newcommand{\comment}[1]{}

\newcommand{\qed}{\nobreak \ifvmode \relax \else
      \ifdim\lastskip<1.5em \hskip-\lastskip
      \hskip1.5em plus0em minus0.5em \fi \nobreak
      \vrule height0.75em width0.5em depth0.25em\fi}

\definecolor{darkspringgreen}{rgb}{0.09, 0.45, 0.27}
\definecolor{forestgreen}{rgb}{0.13, 0.55, 0.13}

\usepackage{array}
\usepackage{comment}
\usepackage{physics}
\usepackage{subcaption}
\usepackage{tikz,tikz-3dplot}
\usepackage[colorlinks=true]{hyperref}
\newcolumntype{C}[1]{>{\centering\let\newline\\\arraybackslash\hspace{0pt}}m{#1}}

\definecolor{yellow2}{rgb}{0.98, 0.80, 0.20}

\definecolor{mygreen}{RGB}{24,174,42}

\title{\centering
Towards Generalized Dimers for GTPs:\\
$\mathcal{N}=2$ Fractional Branes at Infinite Coupling}

\author[a,b,c]{Sebasti\'an Franco,}

\author[d,e]{Diego Rodr\'iguez-G\'omez}

\affiliation[a]{Physics Department, The City College of the CUNY\\
	160 Convent Avenue, New York, NY 10031, USA}
\affiliation[b]{Physics Program and \textsuperscript{$c$}Initiative for the Theoretical Sciences\\
	The Graduate School and University Center, The City University of New York\\
	365 Fifth Avenue, New York NY 10016, USA}
\affiliation[d]{Department of Physics, Universidad de Oviedo \\  
C/ Federico Garc\'ia Lorca  18, 33007  Oviedo, Spain}
\affiliation[e]{Instituto Universitario de Ciencias y Tecnolog\'ias Espaciales de Asturias (ICTEA) \\
 C/~de la Independencia 13, 33004 Oviedo, Spain.}

\emailAdd{sfranco@ccny.cuny.edu}
\emailAdd{d.rodriguez.gomez@uniovi.es}

\abstract{
Generalized Toric Polygons (GTPs) extend the geometric realization of $5d$ superconformal field theories beyond toric Calabi-Yau 3-folds to general $(p,q)$ 5-brane webs ending on 7-branes. We take significant steps towards the generalization of brane tilings for GTPs, or equivalently, the corresponding quiver theories. We focus on GTPs connected to ordinary toric diagrams by polytope mutations. Parallel 5-brane legs of a $(p,q)$ web define $\mathcal{N}=2$ fractional branes bounded by the corresponding parallel zig-zag paths in the brane tiling obtained by treating the GTP as an ordinary toric diagram. We propose that terminating multiple 5-branes on a common 7-brane, the defining feature of GTPs, translates into bringing these zig-zag paths together, thereby shrinking the corresponding $\mathcal{N}=2$ fractional branes to zero size in a process we call $\mathcal{N}=2$ strip condensation. We show that strip condensation follows from mirror symmetry when the coefficients in the Newton polynomial are tuned to the GTP point. We further support this proposal through several non-trivial consistency checks. In particular, it correctly reproduces the expected number of gauge groups, given equivalently by that of the mutation-related toric diagram or by the number of $T$-cones in a tessellation of the GTP. We verify this for all examples previously considered in the literature, as well as for a new infinite family of GTPs with arbitrarily large $T$-cones. Strip condensation drives the corresponding gauge groups to infinite coupling. Confinement then yields quivers that coincide with those of the mutation-related toric diagrams up to vector pairs and adjoint fields, suggesting that GTP quivers are related to those of the corresponding toric diagrams by relevant deformations, extending to GTPs the known correspondence between polytope mutations and relevant deformations.
}

\begin{document}

\maketitle

\section{Introduction}

Quantum Field Theories in $5d$ are notoriously difficult to construct. In particular, while the existence of consistent interacting non-supersymmetric theories remains unclear, supersymmetric theories provide a much better understood landscape. Using the tools of String and M-theory, it is possible to engineer large classes of interacting $5d$ superconformal field theories (SCFTs), which often exhibit remarkable properties such as exceptional global symmetries, non-trivial UV fixed points, and strong coupling phenomena without conventional Lagrangian descriptions. Furthermore, following the modern paradigm, these theories provide powerful insights into lower-dimensional quantum field theories and dualities upon compactification. 

Various approaches for engineering $5d$ SCFTs are known within String and M-theory. The two most salient ones are their geometric engineering via M-theory compactified on singular local Calabi-Yau (CY) 3-folds \cite{Morrison:1996xf,Douglas:1996xp,Intriligator:1997pq} (see also \cite{DelZotto:2017pti,Xie:2017pfl,Alexandrov:2017mgi,Jefferson:2017ahm,Jefferson:2018irk,Bhardwaj:2018yhy,Bhardwaj:2018vuu,Apruzzi:2018nre,Closset:2018bjz,Bhardwaj:2019jtr,Apruzzi:2019vpe,Apruzzi:2019opn,Apruzzi:2019enx,Bhardwaj:2019xeg,Saxena:2020ltf,Apruzzi:2019kgb,DeMarco:2023irn} for recent work), and their realization in terms of webs of $(p,q)$ 5-branes in Type IIB string theory \cite{Aharony:1997ju,Aharony:1997bh} (see also \cite{Benini:2009gi,Bergman:2012kr,Bergman:2013aca,DeWolfe:1998eu,Feng:2005gw,DeWolfe:1999hj}).

From the 5-brane web perspective, it is possible to make properties of $5d$ SCFTs such as the full Higgs manifest by allowing the webs to end on 7-branes. In this case, it becomes possible for multiple 5-branes to terminate on the same 7-brane. Generalized toric polygons (GTPs) were introduced in \cite{Benini:2009gi} as combinatorial tools to describe such configurations and study the constraints imposed by the $s$-rule. GTPs resemble ordinary toric diagrams but contain both black and white dots. White dots on the boundary of the polygon, separating parallel edges, indicate that the corresponding parallel 5-brane legs terminate on a common 7-brane. 

In recent years, significant efforts have been devoted to developing a deeper understanding of GTPs. Their connection to the better understood brane web scenario and its evident similarity to toric geometry suggests there is a rich story yet to be uncovered. Considerable progress has been made on several fronts, including, among others, a better understanding of the geometry of GTPs and the connection between Hanany-Witten (HW) transitions and polytope mutations \cite{Franco:2023flw,Arias-Tamargo:2024fjt, Bourget:2023wlb,Collinucci:2026kom}, the introduction of a new class of quivers, known as twin quivers, which capture various aspects of these constructions such as mutations \cite{Franco:2023flw,Franco:2023mkw}, consistent tessellations of GTPs and connections to the mathematics of $T$-cones \cite{CarrenoBolla:2024fxy,CarrenoBolla:2025rkv}, as well as the study of their associated cluster integrable systems through birational transformations relating them to those for toric diagrams \cite{Kho:2026zwc}. 

Arguably the most compelling and elusive open question concerning GTPs is the identification of their corresponding BPS quivers. The fact that GTPs obviously appear to generalize toric geometry suggests an even more ambitious goal: does a generalization of brane tilings for GTPs exist?\footnote{As a starting point, by a generalization of brane tilings, we mean knowledge of the corresponding quivers and their superpotentials. Whether these theories can also be described by a new class of objects with rich combinatorial and geometric properties is another highly intriguing question.} Such generalized brane tilings would admit, at least, two independent interpretations: as BPS quivers of the $5d$ theories compactified on a circle, and as worldvolume theories on D3-branes probing the corresponding CY 3-folds. Given the far-reaching applications of brane tilings, we expect that a generalization of this sort would open entirely new directions of research.

In this paper, we will take the first steps towards the identification of the generalized brane tilings for GTPs. To do so, we will take a multipronged approach based on reasonable assumptions and will show that they combine into a consistent picture. We believe that the strategies introduced to address this puzzle are, on their own, as interesting as the final proposal. 

This paper is organized as follows. Section \sref{section_background} briefly reviews background material necessary for our presentation, including the brane and geometric engineering of $5d$ SCFTs, brane tilings, and polytope mutations. We also discuss an idea that plays a central role in this paper: the correspondence between the strips bounded by parallel zig-zag paths in brane tilings and $\mathcal{N}=2$ fractional branes. Section \sref{section_general_proposal} presents our general proposal and the various tests we will use to test its consistency. In Section \sref{section_parallel_zigzas_N2_branes} we study how the condensation of $\mathcal{N}=2$ fractional branes accounts for the expected number of gauge groups in the final theory. In Section \sref{section_mirror_symmetry} we discuss how mirror symmetry implies and continuously captures the condensation of $\mathcal{N}=2$ strips. Section \sref{section_confinement} studies the correspondence between strip condensation and confinement in the gauge theory. In Section \sref{section_conclusions}, we conclude and present directions for future research.

\section{Background}

\label{section_background}

For completeness, we briefly review some of the main concepts that will be relevant for our presentation. We refer the reader to the references for further details.

\subsection{Brane and Geometric Engineering of 5d SCFTs}

Two prominent approaches for realizing $5d$ SCFTs are brane webs in Type IIB string theory and geometric engineering in M-theory. Below, we present a brief discussion of the most salient aspects of each of them.

\subsubsection{Brane Webs}

Type IIB string theory contains $(p,q)$ 5-branes. We consider configurations in which these branes share five spacetime dimensions, supporting a $5d$ QFT, while exhibiting a non-trivial structure in the transverse $\mathbb{R}^2$. In a suitable convention, each brane spans a line of slope $(p,q)$ in this transverse plane. These branes can meet supersymmetrically in $\mathbb{R}^2$, provided the $(p,q)$ charges are conserved at each junction, thereby forming a web, dubbed a {\it $(p,q)$ web}, in that plane. When all compact faces of the web shrink to zero size, the low-energy theory on the common $5d$ worldvolume flows to a $5d$ SCFT. These brane configurations were introduced in \cite{Aharony:1997ju,Aharony:1997bh}.

A $(p,q)$ web can be suspended from 7-branes, which are pointlike on the plane of the web, such that $(p_i,q_i)$ 5-branes end on $[p_i,q_i]$ 7-branes. As usual, a branch cut for the axio-dilaton emanates from each 7-brane. When the branch cut of a $[p_j,q_j]$ 7-brane sweeps across a $[p_i,q_i]$ 7-brane in the counterclockwise direction, the latter is transformed according to
\beq
[p'_i,q'_i]^T = M_{[p_j,q_j]} [p_i,q_i]^T \, ,
\eeq
with
\begin{equation}
M_{[p_j,q_j]}=\left(\begin{array}{cc} 1-p_jq_j & p_j^2\\-q_j^2 & 1+p_jq_j\end{array}\right)\,.
\end{equation}
We implicitly assume a standard presentation of the brane, where the branch cuts are oriented radially outgoing along the prongs of the web dictated by the external legs.

A distinctive feature of the $(p,q)$ web realization of $5d$ SCFTs is that the web associated with a given theory is far from unique. To begin with, two webs whose external legs are related by an overall $SL(2,\mathbb{Z})$ transformation are equivalent. More interestingly, seemingly different webs can also be related through crossings involving 7-branes. Indeed, the positions of the 7-branes along the prongs of the external legs are not physical parameters of the $5d$ SCFT. One may therefore move a 7-brane across a junction to the opposite side along its prong. To restore the standard presentation of the web, the monodromy branch cut of the moved 7-brane must be swept across part of the web, inducing the transformations described above. This process typically leads to the creation or annihilation of 5-branes via a Hanany-Witten (HW) transition \cite{Hanany:1996ie}. As a result, two webs that look different can nevertheless describe the same $5d$ theory. In the following, we will generically refer to such transformations as HW transitions.

Supersymmetry of the web suspended from 7-branes imposes severe constraints through the $s$-rule.\footnote{The $s$-rule was originally introduced in \cite{Hanany:1996ie} for D3-branes suspended between D5- and NS5-branes, and it was later generalized to other contexts. In \cite{Benini:2009gi}, it was formulated for $(p,q)$ 5-branes ending on $[r,s]$ 7-branes. In More recently, it was studied in \cite{Bergman:2020myx} using the fact that the SUSY condition for 5-branes ending on 7-branes coincides with that of $(p,q)$ strings ending on $[r,s]$ 7-branes \cite{Iqbal:1998xb}.}  On one hand, it restricts the number of 5-branes that can terminate on the same 7-brane. On the other hand, it constrains the possible deformations of the web into its Coulomb branch.

\begin{figure}[ht!]
\centering
\includegraphics[width=9cm]{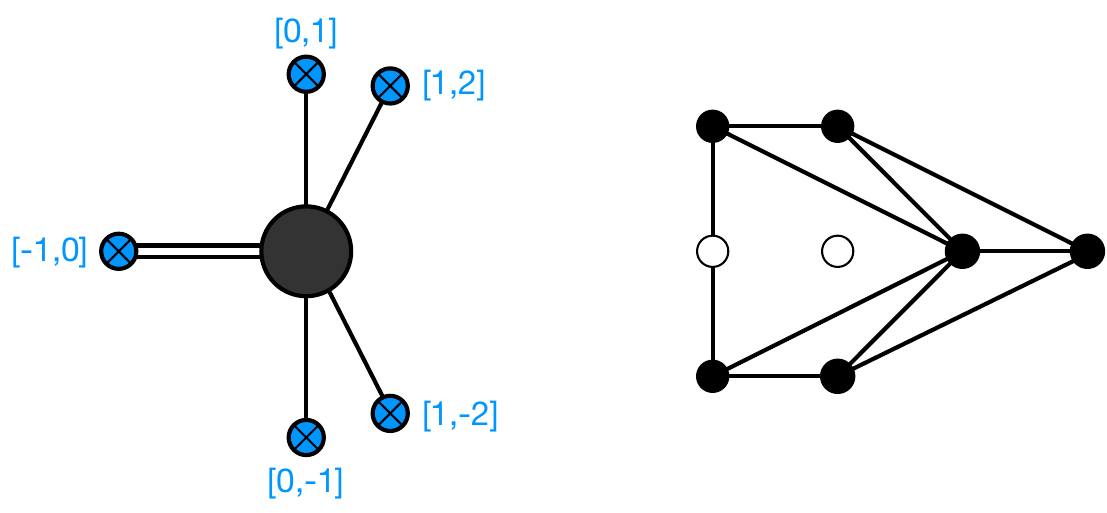}
\caption{A $(p,q)$ web ending on 7-branes and the corresponding GTP.}
\label{example_web_GTP}
\end{figure}

A useful perspective for keeping track of the effects of the 7-branes is through diagram dual to the $(p,q)$ web. To construct this graph, it is convenient to first consider that every external leg can be separated, regardless of whether they end on a common 7-brane or not. The dual graph is then a polytope with integral vertices, represented by black dots in $\mathbb{Z}^2$. The original web is encoded by coloring in white the dots separating the external segments corresponding to 5-branes ending on the same 7-brane.\footnote{Some of the internal points must also be colored white, according to tessellation rules connected to the $s$-rule \cite{CarrenoBolla:2024fxy}.} The resulting polytope is called a generalized toric polygon (GTP). \fref{example_web_GTP} shows an example of a brane web and the corresponding GTP. It was recently proposed that the constraints imposed by the $s$-rule can be implemented by tessellating the GTP into a set of special building blocks known as $T$-cones \cite{Shepherd1988}.\footnote{Strictly speaking, the tessellation may also contain additional building blocks known as locked superpositions \cite{CarrenoBolla:2024fxy}. The tessellation of GTPs with $T$-cones and locked superpositions is closely related, albeit not identical, to previous proposals for tessellations satisfying the $s$-rule \cite{Benini:2009gi}.}

\subsubsection{Geometric Engineering}

Alternatively, $5d$ SCFTs can be geometrically engineered via M-theory on Calabi-Yau 3-fold (CY$_3$) singularities \cite{Morrison:1996xf,Douglas:1996xp,Intriligator:1997pq}. When the CY$_3$ is toric, the standard duality between M-theory and Type IIB string theory relates this construction to $(p,q)$ webs in which all legs end on independent 7-branes. In this case, the 7-branes do not impose any constraints and can effectively be ignored. The $(p,q)$ web then coincides with the toric skeleton of the CY$_3$ that engineers the theory in M-theory \cite{Leung:1997tw}. In terms of the GTP, this corresponds to the absence of white dots, so the GTP reduces to the ordinary toric diagram of the CY$_3$, hence the name GTP in the more general case.

Extending this construction to generic GTPs is an active area of research. It has been recently shown that geometric engineering associated with a generic GTP can be described by the same toric CY$_3$ obtained by interpreting the GTP as an ordinary toric diagram, supplemented by an appropriate freezing of a subset of its deformations \cite{Arias-Tamargo:2024fjt}. Thus, whether $X$ is an ordinary CY$_3$ or a more exotic version with suitably frozen deformations, we may regard any theory encoded by a GTP as being geometrically engineered by M-theory on $X$.

Upon compactifying the $5d$ SCFT on a circle, the resulting $4d$ KK theory is geometrically engineered by Type IIA string theory on $X$. The BPS states of this theory correspond to D0/D4/D6-brane bound states wrapping cycles in $X$. These bound states are captured by the fractional brane quiver associated with $X$, which therefore coincides with the BPS quiver of the $4d$ KK theory. Alternatively, one may consider $N$ Type IIB D3-branes placed at the tip of $X$. In that case, the same fractional brane quiver captures the $4d$ SCFT living on the worldvolume of the D3-branes (see e.g. \cite{Klebanov:1998hh,Morrison:1998cs,Benvenuti:2004dy}).

\subsection{Brane Tilings}

\label{section_brane_tilings}

The purpose of this paper is to develop a generalization of brane tilings associated with GTPs. Since our discussion will often involve ordinary brane tilings as intermediate steps, here we quickly review them. 

The infinite class of $4d$ $\mathcal{N} = 1$ gauge theories living on the worldvolume of D3-branes probing toric CY$_3$ singularities is fully encoded by bipartite graphs on $\mathbb{T}^2$ denoted {\it brane tilings} (or dimer models) \cite{Hanany:2005ve,Franco:2005rj}.\footnote{More precisely, the term brane tiling is commonly used for both the bipartite graph and for the full brane configuration it encodes. In this paper, we will not distinguish between the two possibilities.} In fact, a brane tiling is a physical brane configuration consisting of an NS5-brane wrapping a holomorphic curve from which D5-branes are suspended, which is connected to the D3-branes probing the CY$_3$ by $T$-duality. The holomorphic curve is the zero locus of the Newton polynomial of the toric diagram.

Brane tilings are related to the corresponding gauge theories by a simple dictionary. Faces, edges and nodes of the tiling correspond to unitary gauge group factors, bifundamental or adjoint chiral fields, and superpotential terms, respectively. The same information is equivalently captured by a {\it periodic quiver} on $\mathbb{T}^2$, which is obtained from the brane tiling by graph dualization. In a periodic quiver, terms in the superpotential are encoded in its minimal plaquettes. For detailed presentation of brane tilings see e.g.
\cite{Franco:2005rj,Kennaway:2007tq,Yamazaki:2008bt} and references therein.

Brane tilings significantly simplify the connection between the gauge theories and the corresponding toric CY$_3$'s in both directions. For example, determining the moduli space of the gauge theory, i.e. the probed geometry, reduces to a combinatorial problem involving {\it perfect matchings}, which correspond to the GLSM fields in the toric construction of the CY$_3$.

The construction of brane tilings has been generalized to bipartite graphs on Riemann surfaces, possibly with boundaries, giving rise to the broader class of Bipartite Field Theories (BFTs) \cite{Franco:2012mm, Xie:2012mr,Franco:2012wv,Heckman:2012jh,Franco:2013pg,Franco:2013ana,Franco:2014nca}. BFTs share many of the combinatorial properties and connections to toric geometry (now in various dimensions) exhibited by brane tilings. They have found applications in diverse areas, ranging from scattering amplitudes and on-shell diagrams to mirror symmetry. In the context of this paper, BFTs are related to twin quivers \cite{Franco:2023flw,Franco:2023mkw}.

\subsubsection*{Zig-Zag Paths} 

\begin{figure}[ht!]
\centering
\includegraphics[width=5cm]{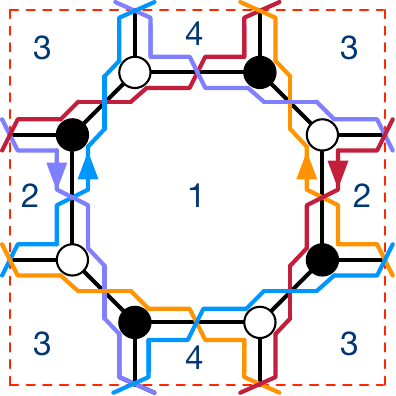}
\caption{Brane tiling for one of the two toric phases of $\mathbb{F}_0$ and its zig-zags.}
\label{tiling_zig_zags_F0_2}
\end{figure}

{\it Zig-zag paths} are fundamental objects in the study of brane tilings and, more generally, bipartite graphs embedded on Riemann surfaces. They are oriented paths on the graph that alternate between making maximal right turns at white nodes and maximal left turns at black nodes \cite{Feng:2005gw,Franco:2012mm}. They can be nicely represented using a double-line notation, in which two zig-zag paths traverse every edge in opposite directions and intersect at its midpoint. \fref{tiling_zig_zags_F0_2} shows a brane tiling with its zig-zag paths.

Like perfect matchings, zig–zag paths are powerful tools for connecting gauge theory and geometry. For example, they underly the so-called Fast Inverse Algorithm for constructing brane tilings from toric diagrams \cite{Hanany:2005ss}. In Section \sref{subsection_infinite_class}, we will see a concrete application that exploits zig-zag paths to identify a class of orbifolds. They have multiple other applications, including the diagnostic of consistency/irreducibility of brane tilings \cite{MR2908565}.

\paragraph{Zig-Zags and Web Legs.}

Every external edge of the toric diagram, or equivalently every leg of the $(p,q)$ web, is in one-to-one correspondence with a zig-zag path of the associated brane tiling. The vector $\eta_i$ of winding numbers of the zig-zag path along the two fundamental cycles of $\mathbb{T}^2$ is equal to the $(p_i,q_i)$ charge of the corresponding leg \cite{Feng:2005gw}. Changing the unit cell of the brane tiling modifies the winding numbers by an $SL(2,\mathbb{Z})$ transformation.  Sides containing multiple edges correspond to parallel legs in the dual $(p,q)$ web. Consequently, the associated brane tiling contains multiple zig-zag paths with identical winding numbers. We will often refer to them as parallel zig-zag paths.

\paragraph{Untwisting and Mirror Symmetry.}

\begin{figure}[ht!]
\centering
\includegraphics[width=9cm]{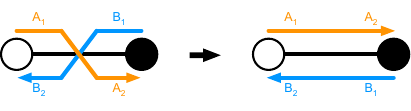}
\caption{The untwisting map.}
\label{untwisting}
\end{figure}

{\it Untwisting} is an operation on zig-zag paths that acts as shown in \fref{untwisting} (see e.g \cite{Feng:2005gw,Franco:2012mm} for further discussion).

Untwisting exchanges zig-zag paths and faces. In doing so, it preserves the bipartite graph but generically changes the Riemann surface on which it is embedded. The bipartite graph on the new surface can be interpreted as a new gauge theory, the {\it twin quiver} \cite{Franco:2023flw,Franco:2023mkw}. Twin quivers encode useful information about polytope mutations, GTPs, the $s$-rule and roots of the Higgs branch.

There is a profound connection between untwisting and mirror symmetry \cite{Feng:2005gw}. When applied to a brane tiling, untwisting produces a bipartite graph embedded on a Riemann surface $\Sigma$, which is the zero locus of the Newton polynomial associated with the toric diagram of the underlying CY$_3$. The surface $\Sigma$ plays a central role in the construction of the mirror geometry \cite{Hori:2000kt,Feng:2005gw} and can be regarded as a fattened version of the corresponding $(p,q)$ web (see Section \sref{section_mirror_symmetry}). More precisely, untwisting maps zig-zag paths to faces surrounding the punctures of $\Sigma$ associated with the corresponding external legs of the brane web.

\subsection{Hanany-Witten Transitions and Polytope Mutations}

\label{section_polytope_mutation}

Geometrically, HW transitions map to {\it polytope mutations} of the dual GTPs \cite{Higashitani:2019vzu}. Let us denote $\eta_i$ the outward pointing vector associated to a side of the polytope. As mentioned earlier, $\eta_i$ is clearly equal to $(p_i,q_i)$, the charges of the associated 5-brane(s).\footnote{Here we are using $\eta_i$ to denote the outward normal to a {\it side} of the polytope, which in principle can involve multiple parallel {\it edges}. We refer the reader to \cite{Franco:2023flw} for details on this nomenclature. We believe that whether we are talking about individual edges or sides will be clear from the context.} We define the intersection number between two sides of the polytope with normal vectors $\eta_{\tilde{i}}$ and $\eta_{\tilde{j}}$ as 
\beq
\langle \eta_{{i}},\eta_{{j}} \rangle = \det \left(\begin{array}{cc} p_{{i}} & q_{{i}} \\ p_{{j}} & q_{{j}} \end{array}\right)
\, .
\label{intersection_sides}
\eeq

The normal vectors satisfy the following relation
\beq
\sum_{{i}} N_{{i}} \eta_{{i}} =0 \, ,
\label{sum_charges}
\eeq
where $N_{{i}}$ is the number of edges in a given side, i.e. the multiplicity of legs in the $(p,q)$-web that have the same $(p_{{i}},q_{{i}})$ charges. Equation \eref{sum_charges} is simply the equilibrium condition for the brane web.\footnote{More precisely, we can consider a vector $\eta_i$ for each set of $N_i$ parallel edges separated by white dots. In particular, there could be more than one such sets in a given side of the GTP. To simplify the presentation, we omit these details in the discussion below. Extending the analysis to include them is straightforward.}

Mutating a polytope with respect to a side ${j}$, corresponds to changing the normal vectors as follows
\beq
\begin{array}{cll}
& \eta_{{j}}'  =  -\eta_{{j}} &  \\[.2cm]
{i} \neq {j}: \ \  & \eta_{{i}}' = \eta_{{i}} + \langle \eta_{{j}},\eta_{{i}} \rangle \eta_j \ \ \ \ & \mbox{for } \langle \eta_{{j}},\eta_{{i}} \rangle > 0 \\[.2cm]
& \eta_{{i}}' = \eta_{{i}} & \mbox{otherwise}
\end{array}
\label{mutation_polytope_1}
\eeq
To satisfy the condition \eref{sum_charges} after the mutation, $N_{{j}}$ must transform according to
\beq
N_{{j}}' = \sum_{{i} \in E_+} N_{{i}} \langle \eta_{{j}},\eta_{{i}} \rangle - N_{{j}}
\, ,
\label{mutation_polytope_2}
\eeq
where $E_{+}$ is the set of sides of the GTP with $\langle \eta_{{j}},\eta_{{i}} \rangle > 0 $. Configurations with $N_{{i}}>1$ must be interpreted as having $N_{{i}}$ 5-branes terminating on the same 7-brane and, therefore, GTPs with white dots.

There is an equivalent transformation in which, for a given mutated side ${j}$ with charge $\eta_{{j}}$, the roles of the $\eta_{{i}}$'s with $\langle \eta_{{j}},\eta_{{i}} \rangle >0$ and $\langle \eta_{{j}},\eta_{{i}} \rangle < 0$ are exchanged in \eref{mutation_polytope_1}. This corresponds to replacing $\langle \eta_{{j}},\eta_{{i}} \rangle$ by $-\langle \eta_{{j}},\eta_{{i}} \rangle$ everywhere in \eqref{mutation_polytope_1}. The two possible mutations correspond to the two ways of moving a 7-brane around the web.

The transformation outlined above is often referred to as {\it combinatorial polytope mutation}. A more refined version, known as {\it algebraic polytope mutation}, implements the transformation at the level of the coefficients of the terms in the Newton polynomial associated with the GTP \cite{GalkinUsnich,akhtar2012minkowski,Arias-Tamargo:2024fjt}. This refined version is essential for understanding the mirror geometry of GTPs \cite{Arias-Tamargo:2024fjt} and the associated cluster integrable systems \cite{Kho:2026zwc}.

\subsubsection*{Brane Tilings from Polytope Mutations}

Reference \cite{Higashitani:2019vzu} introduced an algorithm for constructing the brane tilings corresponding to mutated polytopes, viewed as toric diagrams.\footnote{It is now understood that, generically, the appropriate interpretation of the mutated polytopes is as GTPs (see  \cite{Franco:2023flw} and this paper).} We refer the interested reader to \cite{Higashitani:2019vzu} for details of the construction and to \cite{Franco:2023flw} for explicit examples closely related to the present work. While there are multiple approaches for finding the brane tiling for a given toric diagram, the special feature of this algorithm is that it connects a specific brane tiling for the initial toric diagram to a new brane tiling determined by the mutation. This becomes important whenever we want to investigate the effect of polytope mutation/HW transitions on tilings starting from different toric phases, as in the $dP_2$ examples studied in Sections \sref{section_examples_from_literature} and \sref{section_confinement}.

\subsection{Parallel Zig-Zag Paths and $\mathcal{N}=2$ Fractional Branes}

\label{subsection_parallel_zig_zags_fractional_branes}

Fractional branes at singularities modify the ranks of the gauge groups in the associated quiver gauge theories. In terms of brane tilings, we can think about them as a change in the ranks in some of its faces. In \cite{Franco:2005zu}, fractional branes were classified according to the infrared dynamics they induce. The three resulting classes are: {\it deformation branes}, which trigger complex deformations in the supergravity dual; $\mathcal{N}=2$ {\it branes}, which give rise to $\mathcal{N}=2$ dynamics along their flat directions; and {\it dynamical supersymmetry breaking branes}, whose dynamics leads to the breaking of supersymmetry.

\begin{figure}[H]
\centering
\includegraphics[width=4cm]{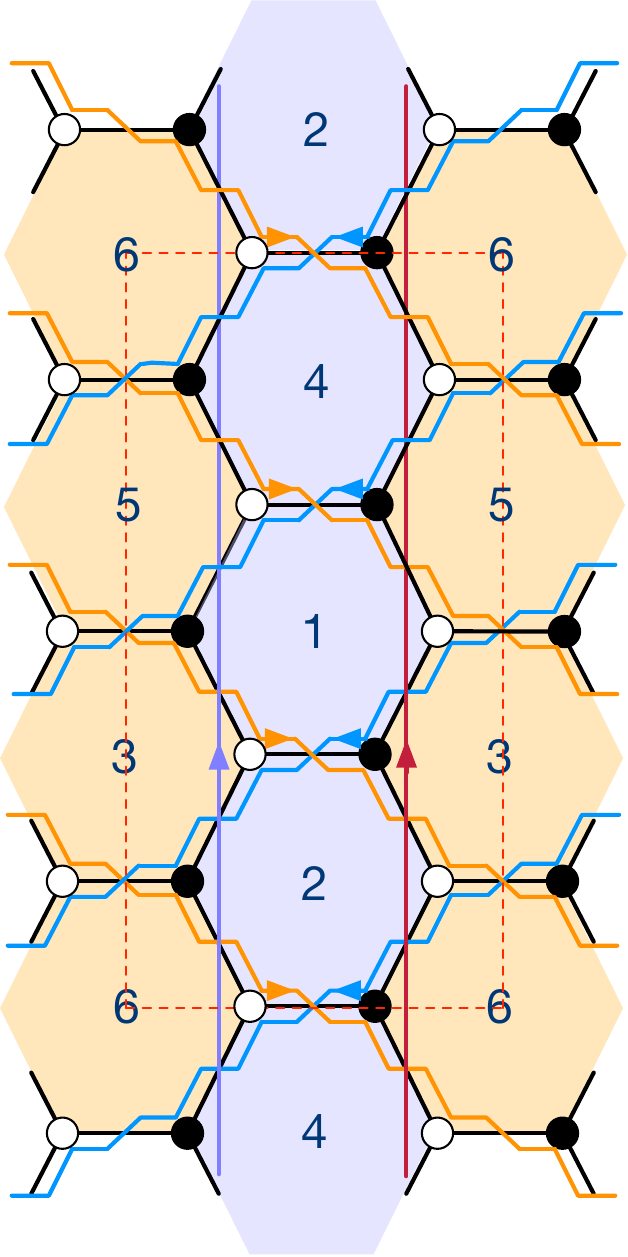}
\caption{Brane tilings for a $\mathbb{C}^3/(\mathbb{Z}_2 \times \mathbb{Z}_3)$ orbifold. The parallel zig-zag paths define two complementary regions, shown in purple and orange. Each of them corresponds to an $\mathcal{N}=2$ fractional brane.}
\label{fractional_branes_C3_Z2_Z3}
\end{figure}

The hallmark of $\mathcal{N}=2$ fractional branes is that their associated faces support a gauge invariant operator that does not appear in the superpotential and therefore parametrizes a flat direction. There is an alternative and very elegant way of characterizing $\mathcal{N}=2$ fractional branes, which is well-known to the experts but is hardly discussed in the literature. They correspond to the faces of a brane tiling lying between two parallel zig-zag paths, as illustrated in \fref{fractional_branes_C3_Z2_Z3} for a $\mathbb{C}^3/(\mathbb{Z}_2 \times \mathbb{Z}_3)$ orbifold. Any pair of parallel zig-zag paths partitions $\mathbb{T}^2$ into two complementary regions, shown in the figure in purple (faces 1, 2 and 3) and orange (faces 4, 5 and 6), each of which corresponds to an $\mathcal{N}=2$ fractional brane.\footnote{In this example, these two sections are basically identical. However, this is not the case in general.}

\section{The General Proposal}

\label{section_general_proposal}

To set the stage for the remainder of the paper, in this section we will provide a brief outline of some of the general ideas we will explore in this paper and explain how they fit together into what we consider to be a compelling picture. 

As reviewed in Section \sref{section_brane_tilings}, zig-zag paths are in one-to-one correspondence with the legs of the $(p,q)$ web. We therefore propose that making two 5-brane legs terminate on the same 7-brane—or, equivalently, introducing a white dot in the GTP between the corresponding edges—amounts to identifying the associated zig-zag paths. While this operation certainly sounds natural, it is important to emphasize that there is further evidence supporting the identification proposal. In \cite{Franco:2023flw}, it was determined that GTPs with white dots correspond to non-toric, Seiberg dual phases of the corresponding twin quivers. Such non-toric phases, meaning both their quivers and superpotentials, are well understood from the basics of Seiberg duality. Remarkably, in \cite{Franco:2023flw} it was also shown that these non-toric twin quivers can be alternatively obtained by starting from the  toric theories obtained from considering the GTP as a toric diagram, building the associated twin quiver by untwisting, and identifying the faces on the untwisted dimer associated to the parallel zig-zags. In the process, which was denoted {\it node merging}, the combination of $n$ $U(1)$ nodes is promoted to a $U(n)$ node. This procedure not only reproduces the quiver but, after some reasonable extension, also generates the correct superpotential.

Once we have established that parallel zig-zags are identified, we should be more explicit about what such identification entails. In Section \sref{section_mirror_symmetry}, we will use mirror symmetry to show that the identification can be understood as starting from the brane tiling for the GTP interpreted as a toric diagram, i.e. without identifications, and continuously bringing the zig-zags together.\footnote{The process for two zig-zag paths and one strip generalizes in the obvious way to $n$ zig-zag paths and the corresponding $n-1$ strips.} In this process, the $\mathcal{N}=2$ fractional brane bounded by these zig-zag paths shrinks to zero size, which corresponds to sending the associated faces to infinite gauge coupling. We will refer to this operation as {\it strip condensation}. In Section \sref{section_confinement}, we will propose that it maps to {\it confinement} of the gauge groups on the strip.

Quiver theories associated with ordinary toric diagrams related by Hanany-Witten transitions (or, equivalently, polytope mutations) are connected by relevant deformations, i.e. by mass terms or higher-order relevant terms in the superpotential \cite{Bianchi:2014qma,Cremonesi:2023psg}. This is only possible if the quivers differ at most by chiral fields forming vector-like pairs or in adjoint representations, which can then be removed by mass terms. Interestingly, we will see that the quivers for GTPs obtained via strip condensation have the precise gauge symmetry and matter content to be connected to the original theories by relevant deformations. This is highly nontrivial and quite compelling, as it suggests that HW transitions also map to relevant deformations of the quiver theories associated with GTPs.

\begin{figure}[H]
\centering
\includegraphics[width=11.3cm]{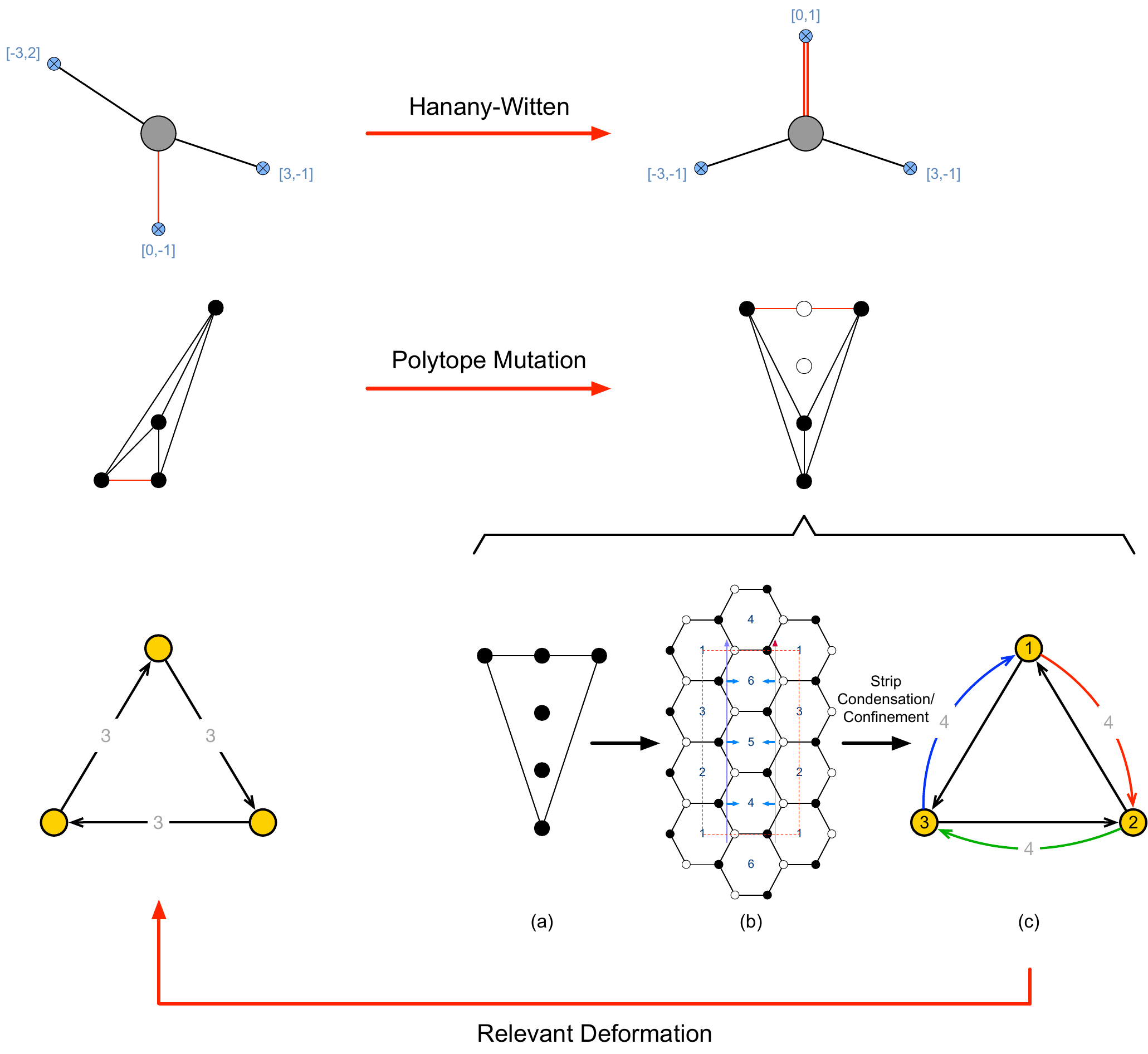}
\caption{Summary of the web of ideas explored in this paper.}
\label{general_idea}
\end{figure}

\fref{general_idea} shows, in an example, the web of ideas that we outlined above and will explore in the rest of the paper. These expectations will be supported by the results we present later in the paper. We regard the fact that all these pieces fit together as evidence in favor of the general picture we propose.

\subsection{Expected Properties: Gauge Groups}

\label{section_expectation_number_of_gauge_groups}

In this section, we discuss our general expectation for the number of gauge groups in the generalized brane tilings associated to GTPs. 

Every gauge group in a quiver describing the worldvolume theory of D3-branes probing a CY$_3$ singularity (or, equivalently, every node in the BPS quiver of the $5d$ theory associated to the CY$_3$) corresponds to a bound state of D-branes wrapping even-dimensional cycles in the geometry. The total number of gauge groups is therefore determined by the number of independent ways in which D-branes (namely D3-, D5-, and D7-branes) can wrap 0-, 2-, and 4-cycles, respectively (D0-, D2- and D4- for the BPS quiver interpretation).

For brane tilings associated with toric diagrams, this number admits a simple combinatorial interpretation in terms of the Euler characteristic of a crepant resolution of the CY$_3$. It is equal to the number of triangles in any triangulation of the corresponding toric diagram. Equivalently, it is given by the area of the toric diagram measured in units in which a fundamental triangle has area 1.

For GTPs, the class of elements in the tessellation of the polygons is extended. In this paper, we will we impose two important restrictions on the GTPs we consider. First, we only consider GTPs that can be obtained from toric diagrams through (sequences of) polytope mutation(s). Second, we require that they admit tessellations composed exclusively of T-cones, namely without the additional building blocks denoted locked superpositions in \cite{CarrenoBolla:2024fxy}. Moreover, we consider what is known in mathematics as a {\it spider triangulation}, in which every $T_n$-cone with $n \geq 2$ has one side lying on a side of the GTP and a vertex at the origin. The remaining portion of the GTP is referred to as the {\it residual singularity} and can be triangulated into minimal triangles in an arbitrary way. 

\begin{figure}[H]
\centering
\includegraphics[width=8cm]{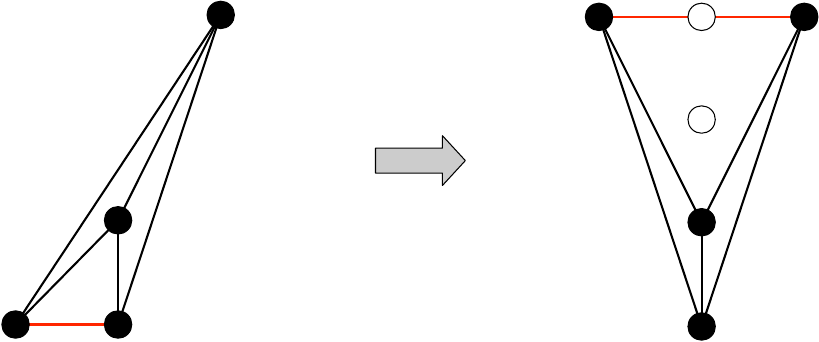}
\caption{Polytope mutation from $dP_0$ to a GTP containing a $T_2$-cone. The mutation acts on the red edge and, in both cases, the origin is the common vertex of all $T$-cones.}
\label{Toric_dP0_mutation_triangulated}
\end{figure}

For the class of theories under consideration, we conjecture:
\begin{itemize}
\item The number of gauge groups in the quiver is equal to the number of T-cones in a tessellation of the GTP.
\end{itemize}
Within this class of theories, this is equal, by construction, to:
\begin{itemize}
\item The number of gauge groups in the quiver is equal to the number of gauge groups in the theory for a toric diagram connected to the GTP by mutations.
\end{itemize}
Remarkably, in Sections \sref{section_parallel_zigzas_N2_branes} and \sref{section_confinement} we will see this counting emerge from a process that, naively, seems completely independent: the confinement of gauge groups in $\mathcal{N}=2$ fractional branes. \fref{Toric_dP0_mutation_triangulated} shows a toric diagram and a GTP connected by mutations and their triangulations/tessellations. We will revisit this example throughout the paper.

This number of gauge groups is also consistent with the counting of variables in the cluster integrable systems associated to GTPs. These integrable systems were recently obtained in \cite{Kho:2026zwc} from the ones associated to toric diagrams via the birational transformations that realize polytope mutations. While extremely interesting, the results of \cite{Kho:2026zwc} left open the question of whether these integrable systems can be derived directly from a suitable generalization of brane tilings, together with a corresponding extension of the construction of Goncharov and Kenyon \cite{Goncharov:2011hp}. The quivers proposed in this paper provide a natural starting point for addressing this question, which we expect to revisit in future work.

\subsection{Preliminary Thoughts on the Number of Gauge Groups}

Let us consider a mutation that transforms an ordinary toric diagram with $G$ minimal triangles into a GTP with $G-1$ minimal triangles and a $T_n$-cone, as shown in \fref{Toric_dP0_mutation_triangulated}. When interpreted as a toric diagram, the $T_n$-cone corresponds to $n^2$ gauge groups. The proposal in Section \sref{section_expectation_number_of_gauge_groups} then implies that these $n^2$ gauge groups are replaced by a single one in the theory associated with the GTP. Consequently, upon reinterpreting the final polytope from a toric diagram to a GTP, we expect to lose $n^2 - 1$ gauge groups. There are two scenarios one could imagine for how this happens:
\begin{enumerate}
\item There are gauge groups in the theory in one-to-one correspondence with minimal triangles inside the $T_n$-cone. All these gauge groups combine into a single one, leading to the reduction by $n^2-1$ gauge groups.
\item There is a process in the brane tiling that shrinks $n^2-1$ gauge groups and make them disappear. 
\end{enumerate}
While the first scenario appears natural, it is important to emphasize that even in the case of toric diagrams, the mapping between minimal triangles in a triangulation and nodes in the quiver is not straightforward. In the following sections, we will present evidence for the strip condensation process discussed in Section \sref{section_general_proposal} and, moreover, will show that it precisely realizes the second scenario.

\section{Parallel Zig-Zag Paths and $\mathcal{N}=2$ Fractional Branes}

\label{section_parallel_zigzas_N2_branes}

In the previous section, we discussed how the $n^2$ gauge groups associated with a $T_n$-cone, when interpreted as a toric diagram, are expected to be replaced by a single gauge group in the corresponding GTP. More appropriately, we expect the number of gauge groups to be reduced by $n^2-1$.

\subsection{A General Conjecture: The Brane Tiling Counterparts of $T$-Cones}

Below, we will show in a plethora of examples that strip condensation indeed leads to the desired reduction in the number of groups. But is there a general mechanism? As we know:
\begin{itemize}
\item Every $T_n$ corresponds to $n$ parallel zig-zags.
\item These $n$ zig-zags divide the brane tiling into $n$ $\mathcal{N}=2$ strips between them.
\end{itemize}

We conjecture that, when a GTP containing a $T_n$-cone is interpreted as a toric diagram, $n-1$ of the $n$ zig-zag strips---namely, those that condense in the GTP limit---have the following structure:
\begin{itemize}
\item Each of the $n-1$ strips contains $n+1$ faces.
\item The total number of faces in these strips is then $(n-1)(n+1)=n^2-1$, which agrees precisely with the necessary counting of gauge groups to be eliminated.
\end{itemize}
This structure can be regarded as the brane tiling counterpart of a $T_n$-cone. The remaining, non-condensing strip can have a different structure, as explicitly shown in some of the examples presented below. 

Below, we present several explicit examples showing how strip condensation realizes the desired reduction, but also satisfy our conjecture. To the best of our knowledge, the evidence presented here encompasses all the examples of toric diagram $\to$ GTP transitions that have been previously studied in the literature in other contexts, including twin quivers \cite{Franco:2023flw}, mirror symmetry \cite{Arias-Tamargo:2024fjt}, and integrable systems \cite{Kho:2026zwc}. We also introduce a new infinite class of models and show that it satisfies our proposal. We believe it could be possible to prove this conjecture using the construction of \cite{Higashitani:2019vzu}. We leave this for future work.

Finally, it is interesting to note that in all the examples we consider, the faces in these strips are hexagons. While our main conclusions regarding the number of gauge groups do not depend on this property, it would be interesting to determine whether this is true in general.

\subsection{Examples from the Literature}

\label{section_examples_from_literature}

We first consider strip condensation in several examples that have previously appeared in the literature in the study of mutations and GTPs. 

\subsubsection*{Example 1: Mutation of $dP_0$}

Let us consider the mutation of $dP_0$ shown in \fref{Toric_dP0_mutation_triangulated}. This mutation was  previously studied in \cite{Arias-Tamargo:2024fjt} and, as shown in the figure, generates a $T_2$-cone.

\begin{figure}[H]
\centering
\includegraphics[width=4cm]{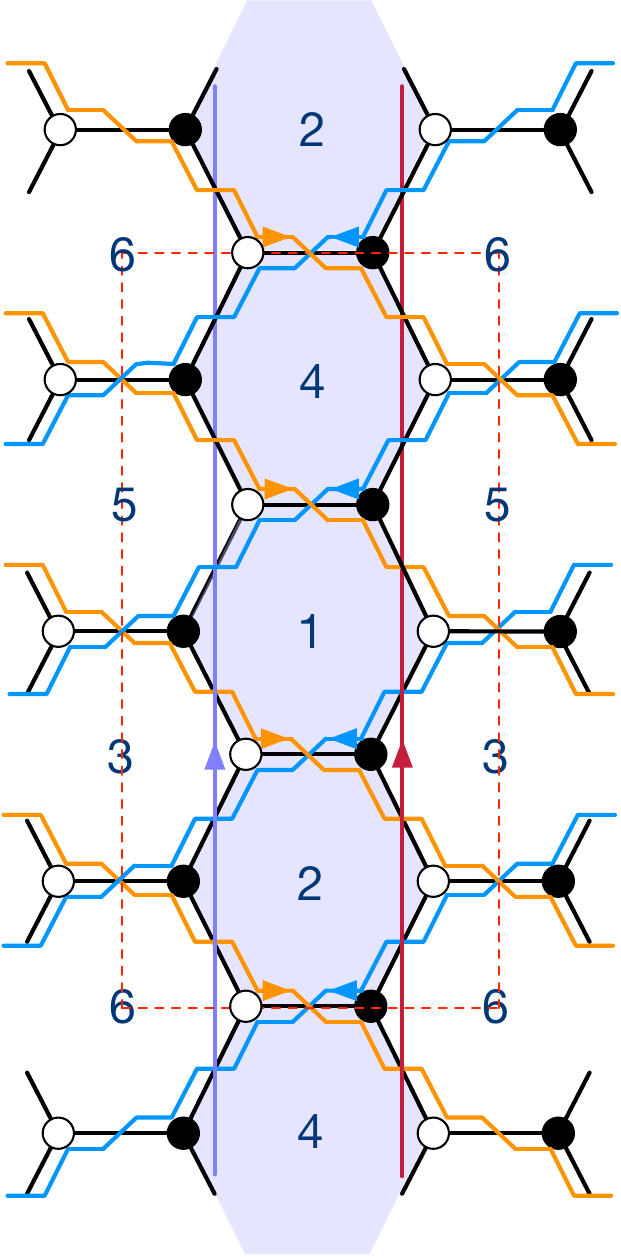}
\caption{Brane tiling for the GTP on the right of \fref{Toric_dP0_mutation_triangulated}, when interpreted as a toric diagram.}
\label{tiling_Z2_Z3}
\end{figure}

When the GTP is considered as a toric diagram, it corresponds to a $\mathbb{C}^3/(\mathbb{Z}_2 \times \mathbb{Z}_3)$ orbifold, whose brane tiling is shown in \fref{tiling_Z2_Z3}.\footnote{We use a unit cell that is equal to the one in \cite{Arias-Tamargo:2024fjt}. This choice gives rise to winding numbers for zig-zag paths that correspond to an $SL(2,\mathbb{Z})$ transformation of the polytope in \fref{Toric_dP0_mutation_triangulated}.} In the figure, we also show the zig-zag paths. In agreement with our conjecture, the two parallel zig-zag paths define a strip containing 3 hexagons, which is indicated in light purple. Shrinking this strip leaves us with 3 gauge groups, as in the original $dP_0$ theory, and as it also follows from the two minimal triangles and one $T_2$-cone in the GTP.

This example is slightly degenerate because, due to the symmetry between the strip considered in \fref{tiling_Z2_Z3} and its complement, as already shown in \fref{fractional_branes_C3_Z2_Z3}. Using mirror symmetry, we will investigate how this degeneracy is lifted in Section \sref{section_mirror_symmetry}.

\subsubsection*{Example 2: Mutation of $dP_1$}

\begin{figure}[H]
\centering
\includegraphics[width=8cm]{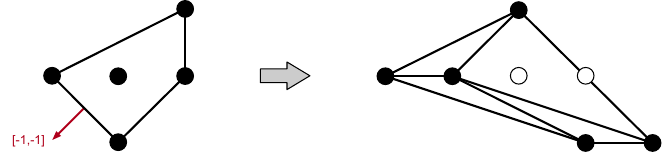}
\caption{Polytope mutation from $dP_1$ to a GTP containing a $T_2$-cone. The red normal vector indicates the mutated edge.}
\label{Toric_dP1}
\end{figure}

Let us now consider the mutation of $dP_1$ shown in \fref{Toric_dP1}, which generates a $T_2$-cone. This mutation was considered in \cite{Franco:2023flw}. 

\begin{figure}[H]
\centering
\includegraphics[width=7cm]{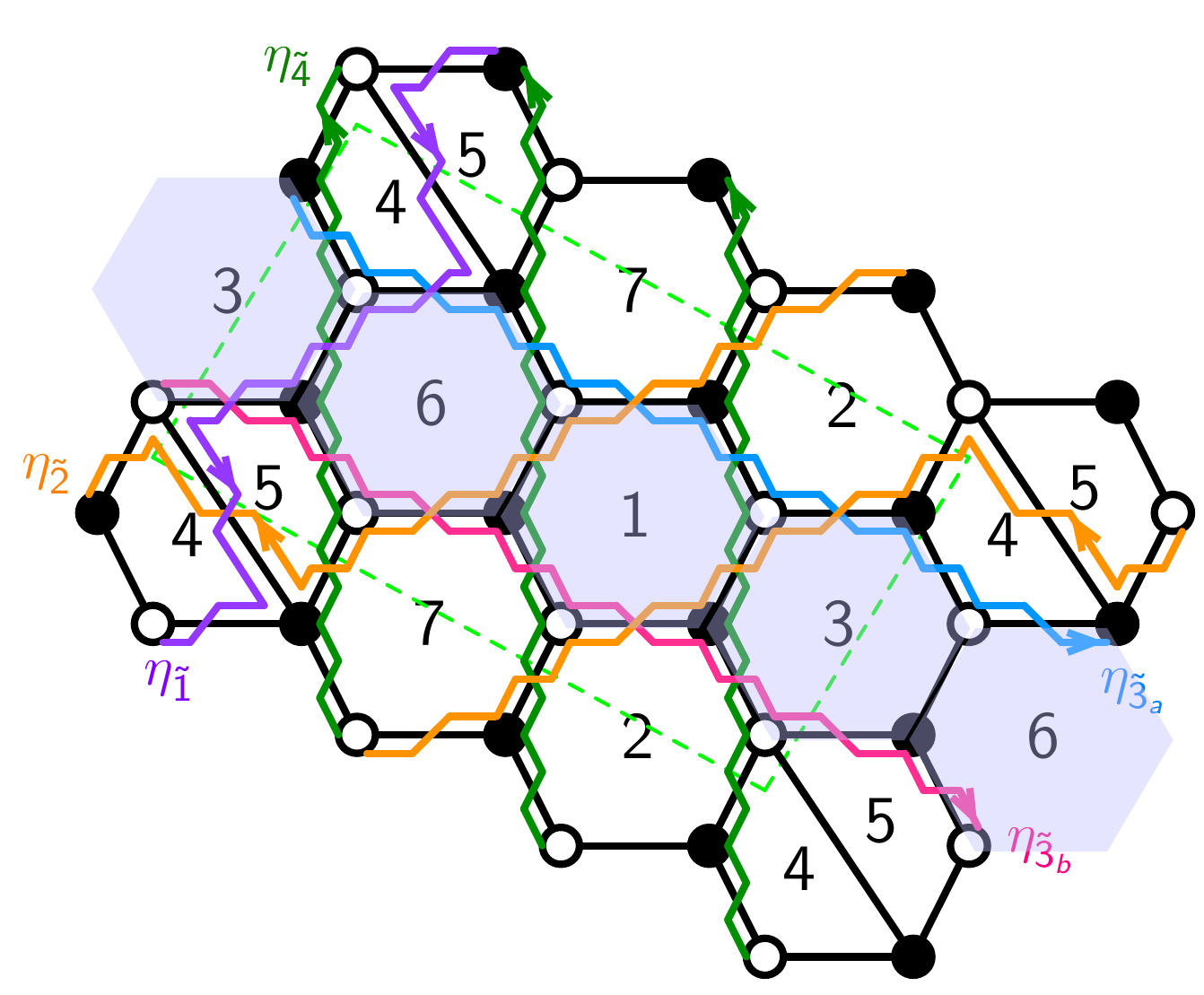}
\caption{Brane tiling for the GTP on the right of \fref{Toric_dP1}, when interpreted as a toric diagram.}
\label{Tiling_mutation_dP1}
\end{figure}

\fref{Tiling_mutation_dP1} shows the brane tiling for a toric phase once the GTP is considered as a toric diagram. The toric diagram corresponds to $X^{3,3}$ in the classification of \cite{Hanany:2005hq}. Once again, we show the strip to be condensed in purple. The strip contains 3 gauge groups, which is indeed the necessary number to go from 7 to 4 gauge groups, as indicated by the tessellation of the GTP in \fref{Toric_dP1}. This example is particularly interesting because the complement of the purple strip, which is itself an $\mathcal{N}=2$ strip, contains a different number of gauge groups, namely 4. The existence of a strip with the right number of gauge groups is already non-trivial. We will argue in Section \sref{section_mirror_symmetry}, that mirror symmetry can be used to single out the condensing strip.

\subsubsection*{Examples 3 and 4: Mutation of $dP_2$}

\begin{figure}[H]
\centering
\includegraphics[width=5.7cm]{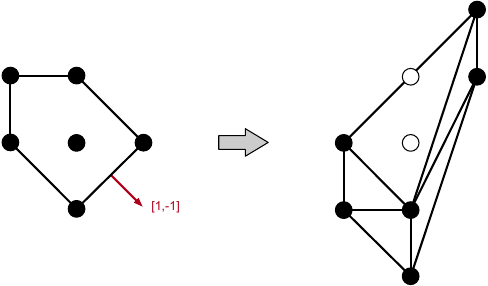}
\caption{Polytope mutation from $dP_2$ to a GTP containing a $T_2$-cone.}
\label{Toric_dP2}
\end{figure}

Let us now consider the mutation of $dP_2$ shown in \fref{Toric_dP2}, which, once again, generates a $T_2$-cone. This mutation was considered in \cite{Franco:2023flw}. 

\begin{figure}[H]
\centering
\includegraphics[width=4.7cm]{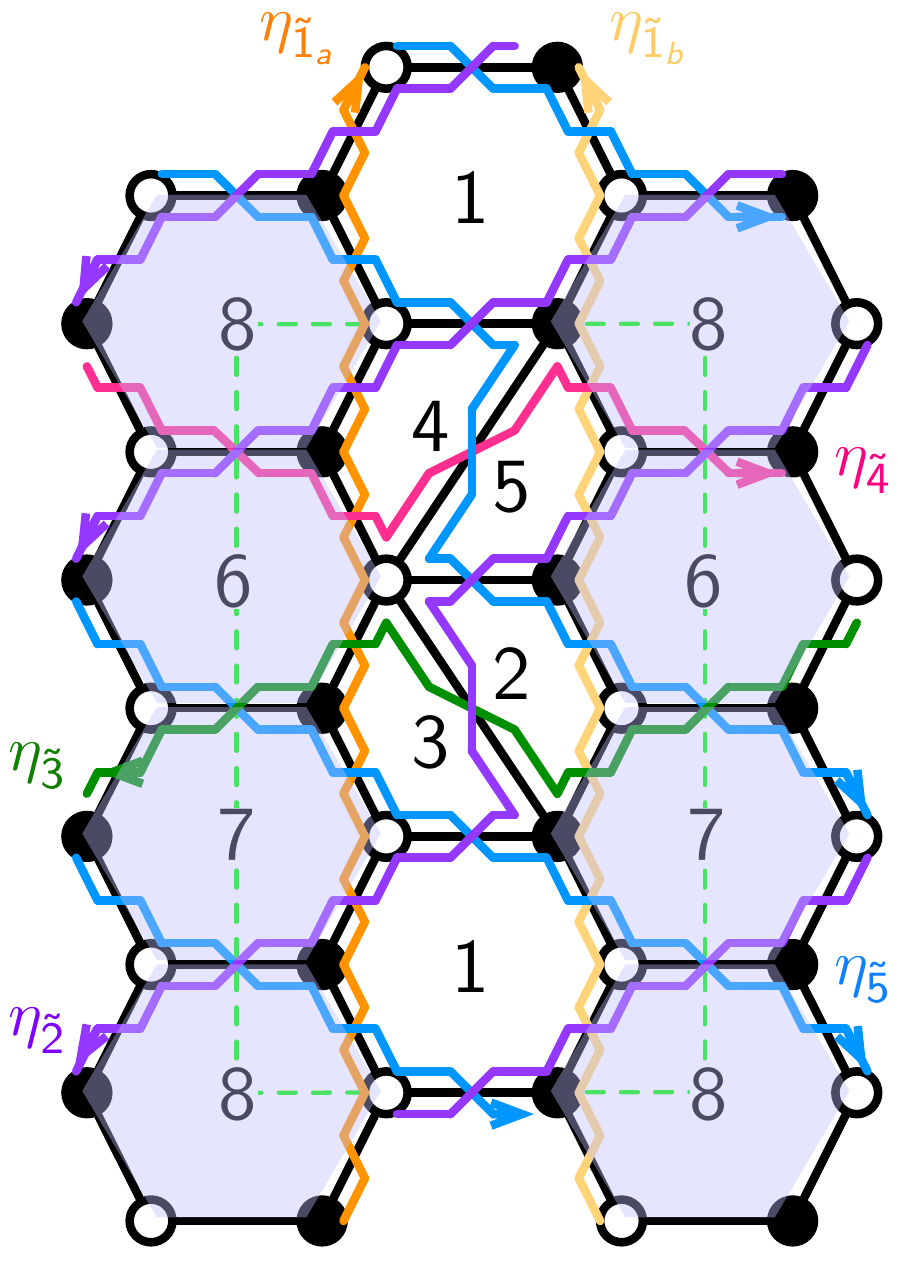}
\caption{Brane tiling for the GTP on the right of \fref{Toric_dP2}, when interpreted as a toric diagram. This brane tiling was obtained from the second toric phase of $dP_2$.}
\label{Tiling_mutation_dP2_1}
\end{figure}

\fref{Tiling_mutation_dP2_1} shows the brane tiling for a toric phase once the GTP is considered as a toric diagram.\footnote{As in the previous example the choice of unit cell gives rise to winding numbers for zig-zag paths that correspond to an $SL(2,\mathbb{Z})$ transformation of the polytope in \fref{Toric_dP2}.} In agreement with our conjecture, the two parallel zig-zag paths define a strip containing 3 hexagons, which is indicated in light purple. Shrinking this strip leaves us with 5 gauge groups, as in the original $dP_2$ theory, and as it also follows from the 4 minimal triangles and 1 $T_2$-cone in the tessellation of the GTP. As in the previous example, the condensing strip is quite different from its complement. This brane tiling was obtained from the what is often called the second toric phase of $dP_2$ \cite{Feng:2002zw,Franco:2005rj} using the algorithm in \cite{Higashitani:2019vzu}.

\begin{figure}[H]
\centering
\includegraphics[width=6.5cm]{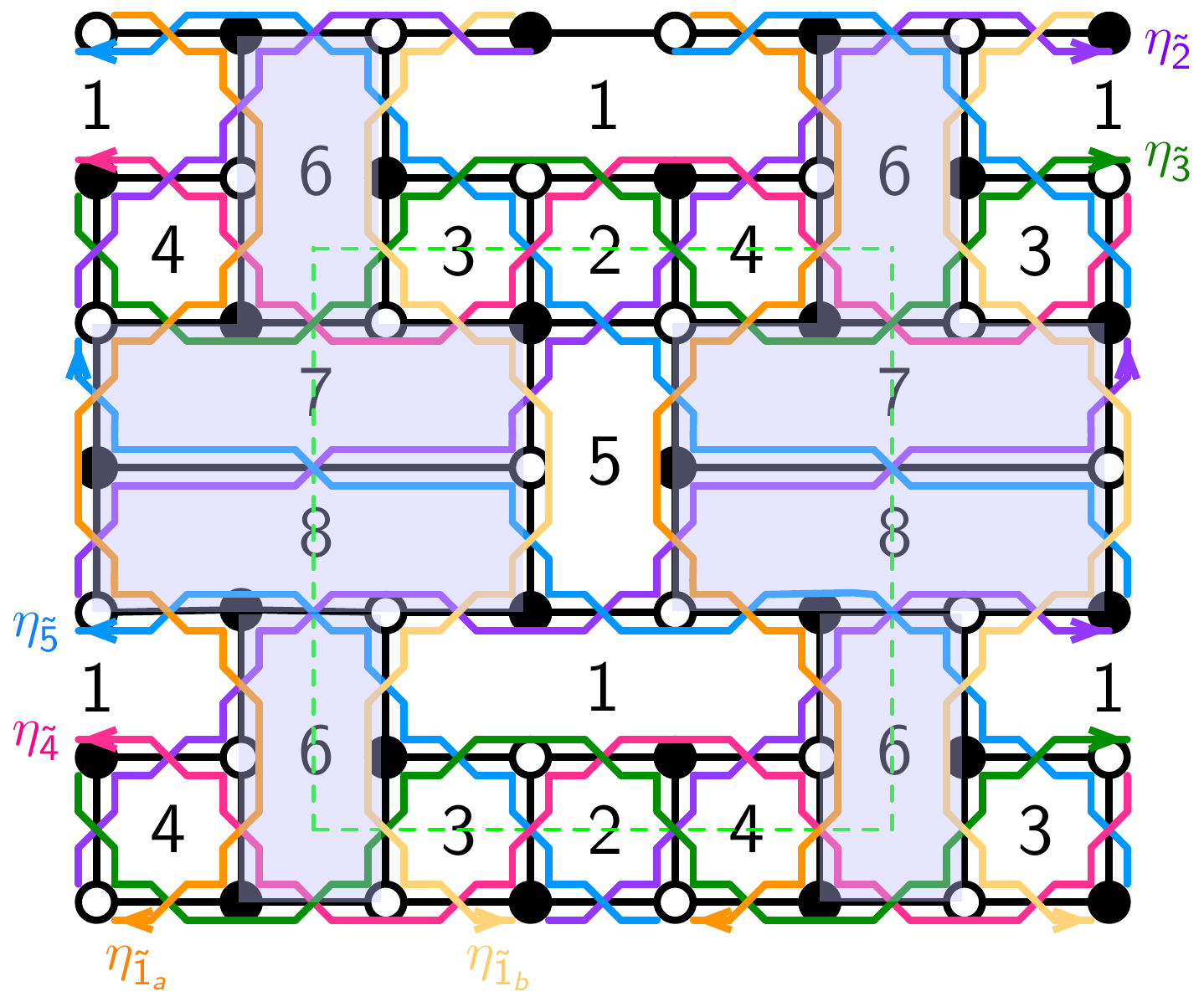}
\caption{Brane tiling for the GTP on the right of \fref{Toric_dP2}, when interpreted as a toric diagram. This brane tiling was obtained from the first toric phase of $dP_2$.}
\label{Tiling_mutation_dP2_2}
\end{figure}

This example is extremely interesting because we can consider how things work in a different toric phase, which we show in \fref{Tiling_mutation_dP2_2}. As explained in \cite{Franco:2023flw}, this brane tiling is obtained by starting instead from the first toric phase of $dP_2$ \cite{Feng:2002zw,Franco:2005rj} and applying the algorithm in \cite{Higashitani:2019vzu}. In this alternative toric phase, the two parallel zig-zag paths also define a strip containing 3 hexagons, in agreement with our expectations.

\subsection{An Infinite Class of GTPs}

\label{subsection_infinite_class}

It would be interesting to test our proposals in a wider set of GTPs. Moreover, it would be desirable to move beyond the previous example and consider mutations that generate $T_n$-cones with $n>2$. In this section we present an infinite family of models that addresses both issues.

\subsubsection*{An Example with a $T_3$-Cone}

\begin{figure}[ht!]
\centering
\includegraphics[width=6cm]{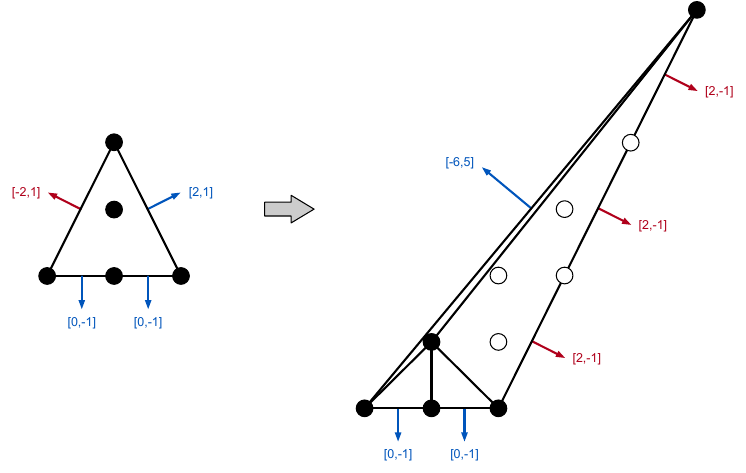}
\caption{Polytope mutation from a toric diagram to a GTP containing a $T_3$-cone.}
\label{Toric_T3}
\end{figure}

Let us first consider the mutation in \fref{Toric_T3}, which generates a $T_3$-cone.\footnote{While the initial theory contains a pair of parallel $[0,1]$ zig-zag paths, they correspond to independent legs of the $(p,q)$ web. The same is true for all examples in this subsection.} 

\begin{figure}[ht!]
\centering
\includegraphics[width=8cm]{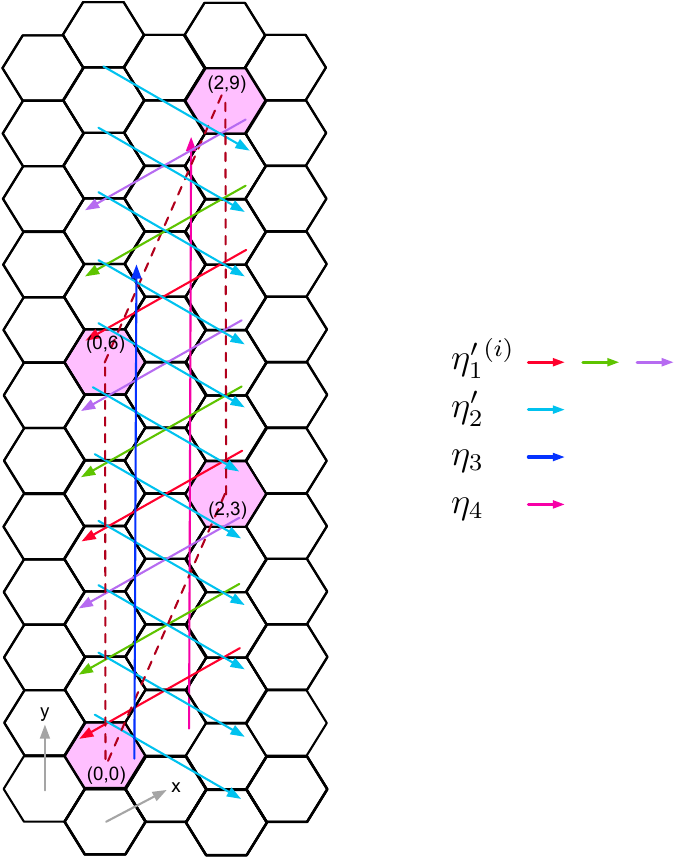}
\caption{Brane tiling for the GTP on the right of \fref{Toric_T3}, when regarded as a toric diagram. It corresponds to a $\mathbb{C}^3/(\mathbb{Z}_2 \times \mathbb{Z}_6)$ orbifold.}
\label{brane_tiling_T3}
\end{figure}

Thought of as a toric diagram, the resulting GTP corresponds to a $\mathbb{C}^3/(\mathbb{Z}_2 \times \mathbb{Z}_6)$ orbifold. The corresponding brane tiling is shown in \fref{brane_tiling_T3}. We have highlighted the faces at the corners of the unit cell in pink. We have also given their coordinates, using the $x$ and $y$ directions indicated in the figure. We will later explain, using the general case for this family, how to determine the brane tiling corresponding to this specific orbifold.

For this family, we have picked the unit cell such that gives rise exactly to the desired winding numbers, instead of an $SL(2,\mathbb{Z})$ transformation of them. We have only flipped the sign of the winding numbers in the vertical direction, which is merely a vertical reflection and results in an equivalent GTP.

\begin{figure}[H]
\centering
\includegraphics[width=2.3cm]{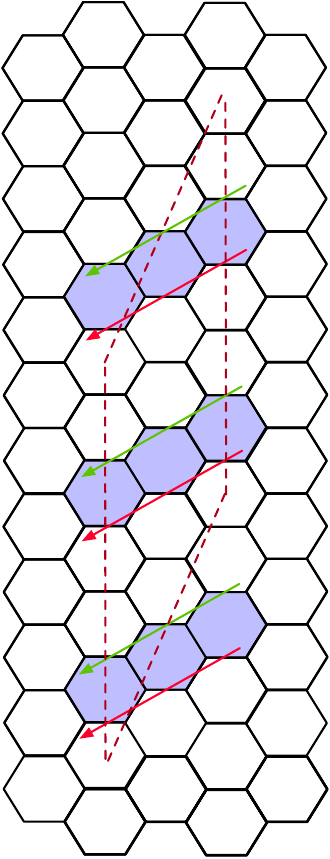}
\caption{An $\mathcal{N}=2$ strip bounded by two parallel zig-zag paths. It contains 4 hexagons.}
\label{strip_brane_tiling_T3}
\end{figure}

\fref{strip_brane_tiling_T3} shows the strip generated by one pair of zig-zag paths (red and green). Given the periodicity of the tiling, we see that the strip contains 4 hexagons. The 3 parallel zig-zags generate 2 strips, which contain 8 hexagons.\footnote{In the family of examples considered in this section, the complement of the strips we consider, is another identical strip.} Once they condense, they leave behind 4 gauge groups, in agreement with the tessellation of the GTP in \fref{Toric_T3}.

\vspace{-.5cm}

\subsubsection*{An Example with a $T_5$-Cone}

\begin{figure}[H]
\centering
\includegraphics[width=6.8cm]{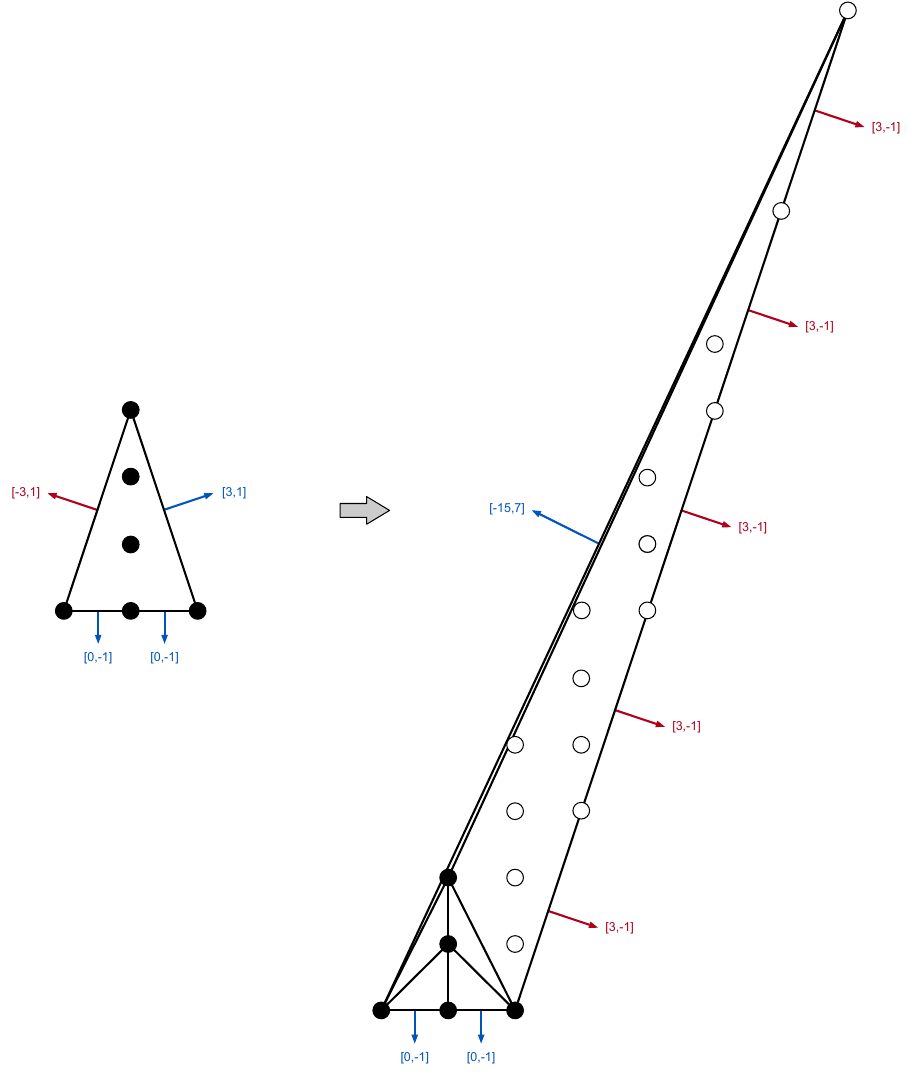}
\caption{Polytope mutation from a toric diagram to a GTP containing a $T_5$-cone.}
\label{Toric_T5}
\end{figure}

Let us now consider the next model in the family, which is shown in \fref{Toric_T5}. The mutation generates a $T_5$-cone. When viewed as a toric diagram, the GTP corresponds to a $\mathbb{C}^3/(\mathbb{Z}_2 \times \mathbb{Z}_{15})$ orbifold.

\vspace{-.2cm}

\begin{figure}[H]
\centering
\includegraphics[width=7cm]{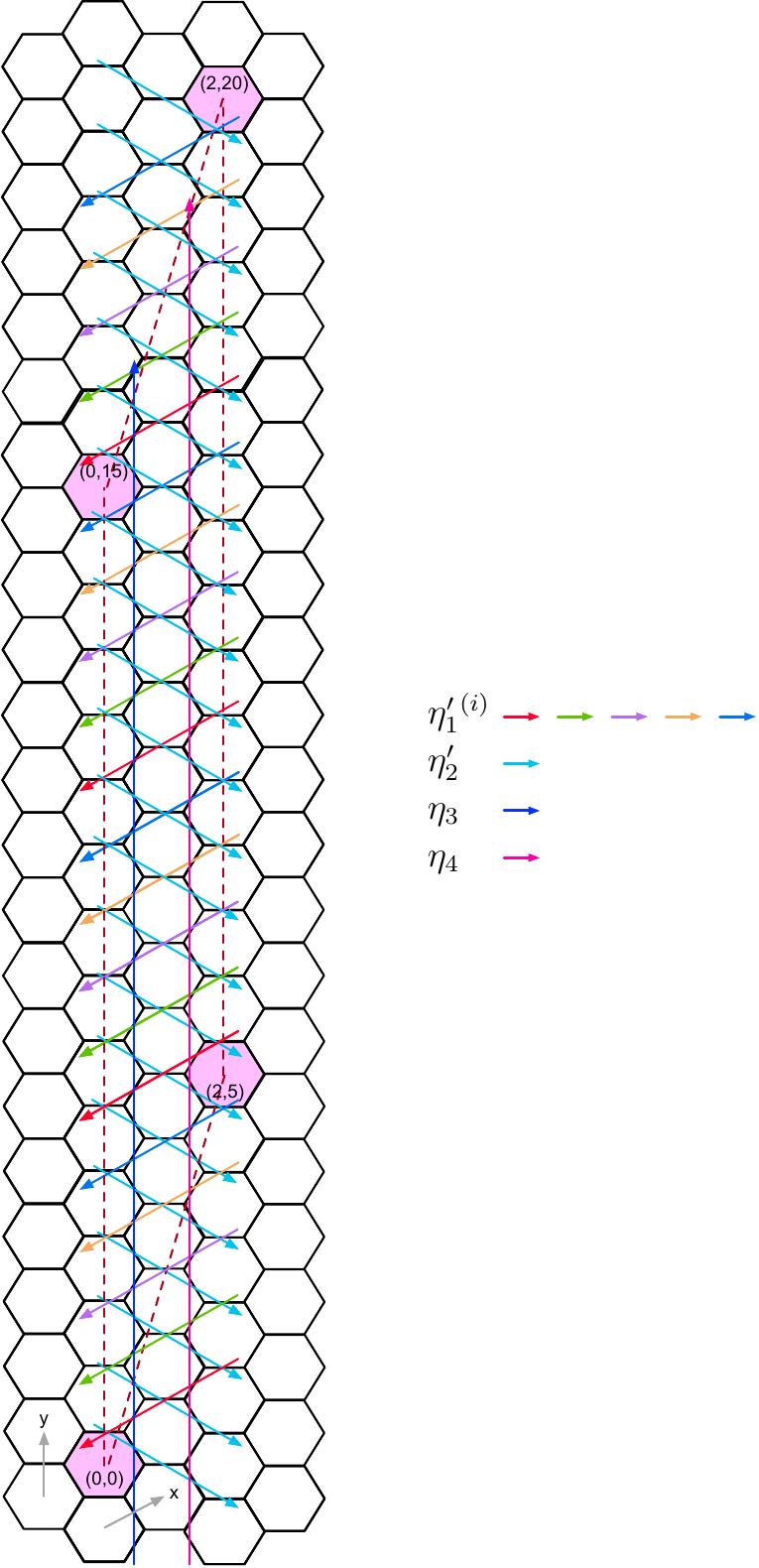}
\caption{Brane tiling for the GTP on the right of \fref{Toric_T5}, when regarded as a toric diagram. It corresponds to a $\mathbb{C}^3/(\mathbb{Z}_2 \times \mathbb{Z}_{15})$ orbifold.}
\label{brane_tiling_T5}
\end{figure}

The corresponding brane tiling is shown in \fref{brane_tiling_T5}.

It is easy to check that the five parallel zig-zags define 4 strips, each of them containing 6 hexagons. After condensing them, we are left with $30-24=6$ faces, as expected (from both the counting of gauge groups in the original theory and the tessellation of the GTP in \fref{brane_tiling_T5}).

\subsubsection*{The General Case}

\label{section_general_Tn_case}

\begin{figure}[h!]
\centering
\includegraphics[width=8.5cm]{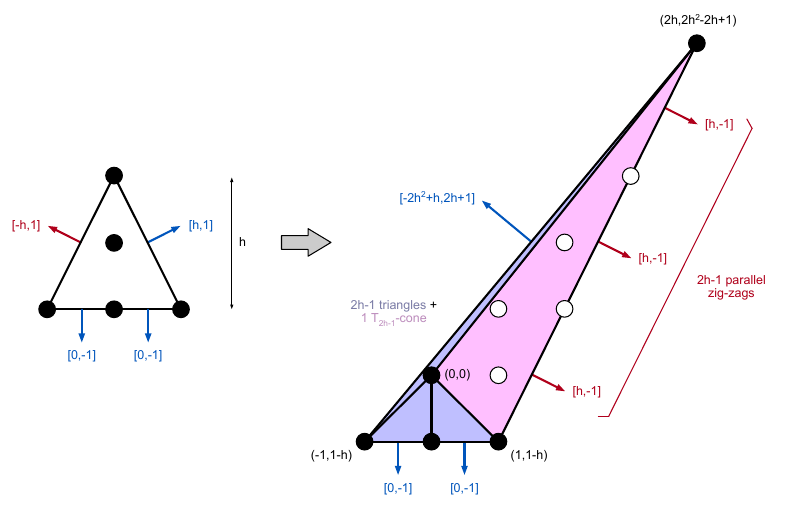}
\caption{An infinite family of polytope mutations between $\mathbb{C}^3/(\mathbb{Z}_2 \times \mathbb{Z}_h)$ orbifolds and GTPs with $T_{2h-1}$-cones.}
\label{Toric_Th}
\end{figure}

We are now ready to consider the general case. \fref{Toric_Th} shows how the mutation works. The original toric diagram is an isosceles triangle with base of length 2 and height $h$, which corresponds to a $\mathbb{C}^3/(\mathbb{Z}_2 \times \mathbb{Z}_h)$ orbifold. This corresponds to a theory with $2h$ gauge groups. We mutate the toric diagram on the $\eta_1=[-h,1]$ edge. The intersection number between this edge and the other non-horizontal side of the triangle, which has $\eta_2=[h,1]$, is $\langle \eta_2,\eta_1 \rangle = 2h$. This means that in the twin quiver we would end with a rank $2h-1$ node. The mutated zig-zag $\eta_1$ turns into $2h-1$ zig-zags $\eta_1^{\prime(i)}=[h,-1]$, $i=1,\ldots,2h-1$. Equivalently, the minimal triangle associated with $\eta_1$ transforms into a $T_{2h-1}$-cone. In summary, the zig-zag content of the mutated theory is
 \beq
 \begin{array}{ccl}
\eta_1^{\prime(i)} & = & [h,-1]  \ \ \ \ \ i=1,\ldots,2h-1 \\
 \eta'_2 & = & [-2h^2+h,2h+1] \\
\eta_3 & = & [0,-1]  \\ 
  \eta_4 & = & [0,-1] 
\end{array}
 \eeq
 We can directly get $\eta'_2$ using the transformation rule for zig-zags under mutation or by using the values and multiplicities of $\eta_1^{\prime(i)}$, $\eta_3$, $\eta_4$ and requiring that all zig-zag paths add up to zero. In \fref{Toric_Th}, we also specify the coordinates of the corners of the GTP, which take into account the position of the origin.
 
\begin{figure}[H]
\centering
\includegraphics[width=8.5cm]{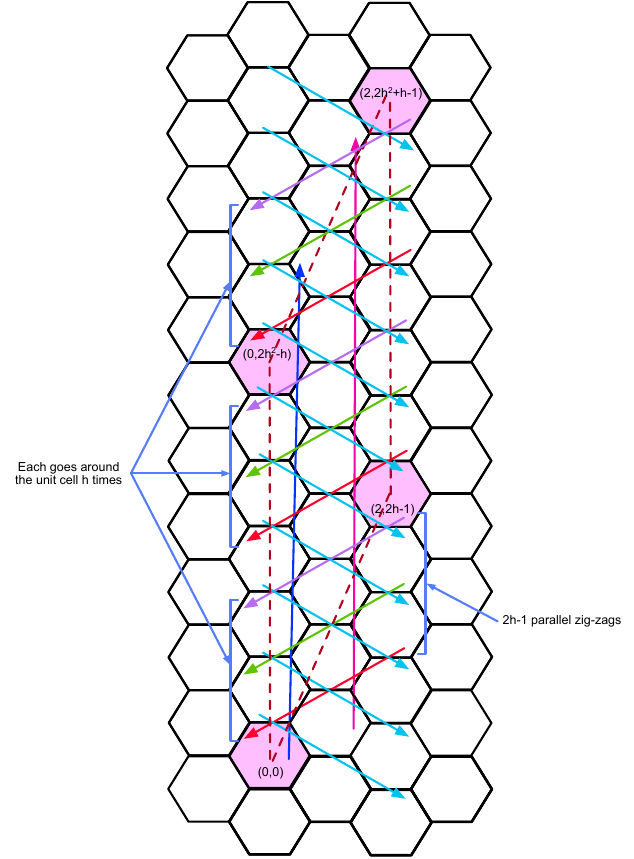}
\caption{Brane tiling for the GTP on the right of \fref{Toric_Th}, when regarded as a toric diagram. It corresponds to a $\mathbb{C}^3/(\mathbb{Z}_2 \times \mathbb{Z}_{2h^2-h})$ orbifold.}
\label{brane_tiling_Th}
\end{figure}

When interpreted as a toric diagram, the final GTP corresponds to a $\mathbb{C}^3/(\mathbb{Z}_2 \times \mathbb{Z}_{2h^2-h})$ orbifold. Determining the action, or equivalently identifying the unit cell of an hexagonal lattice, that gives rise to this specific product orbifold is, in general, a nontrivial problem.\footnote{See e.g. \cite{Hanany:2010cx,Davey:2010px,Hanany:2010ne,Davey:2011dd} for general discussions of abelian orbifolds of $\mathbb{C}^3$. This example, however, goes well beyond the cases considered in these references.} We now explain an elegant prescription to do so. \fref{brane_tiling_Th} shows the general form of the brane tiling. 

Regardless of the unit cell, the zig-zag paths go in three possible directions: -30$^\circ$, -120$^\circ$ and 90$^\circ$. Moreover, without loss of generality, we pick the following correspondence between the winding numbers of zig-zags and their directions:
\beq
\begin{array}{ccccr}
\eta_3, \eta_4 & : & [0,-1] & \to & 90^\circ \\
\eta_1^{\prime(i)} & : & 2h-1 \times [h,-1] & \to & -120^\circ \\
\eta'_2 & : & \left[-2h^2+h,2h+1\right] & \to & -30^\circ
\end{array}
\eeq
Note that, es in the previous examples in this class, we have flipped the vertical winding numbers in the figure. The unit cell is determined by demanding the two first sets of zig-zag paths to have the correct winding numbers. Once this is achieved, $\eta'_2$ will have the correct winding numbers by construction. To achieve this, we proceed as follows:
\begin{itemize}
\item Without loss of generality, we can pick two opposite sides of the unit cell to be vertical. This will produce the correct winding numbers for $\eta_3$ and $\eta_4$. This also fixes the width of the unit cell to be 2, in agreement with the $\mathbb{Z}_2$ part of the orbifold group.
\item The diagonal side of the unit cell (and its opposite, parallel side) is determined by requiring that each of the $2h-1$ $\eta_1^{\prime(i)}$ zig-zags cross it exactly once. 
\item The height of the unit cell is $2h^2-h$. This follows from the $\mathbb{Z}_{2h^2-h}$ part of the orbifold group. More fundamentally, it follows from the fact that, to achieve their $[h,-1]$ winding numbers,  each $\eta_1^{\prime(i)}$ must intersect the vertical boundary of the unit cell $h$ times. This requires stacking $h$ ``copies" of the $2h-1$ $\eta_1^{\prime(i)}$ zig-zags vertically, as shown in \fref{brane_tiling_Th}, leading to the $2h^2-h$ height.
\end{itemize}

\paragraph{Strip Counting.} In the general case, we then have $2h-1$ parallel $\eta_1^{\prime(i)}$ zig-zags. They define $2h-2$ strips (there is, obviously, an additional strip in the complement). Each of the strips contains $2h$ hexagons. Aa result, there are $(2h-2)2h=4h^2-2h$ hexagons in these strips, which we claim will disappear under condensation. In agreement with our proposal, this is 1 less than the area of a $T_{2h-1}$-cone in terms of minimal triangles, which is $4h^2-2h+1$. After condensation, the number of surviving gauge groups is $2h$, equal to the number of gauge groups in the complement of the condensed strips. This coincides precisely with the area of the original toric diagram in \fref{Toric_Th}, measured in units of minimal triangles, which is itself equal to the number of gauge groups in the brane tiling prior to mutation. We conclude that everything works beautifully for this infinite class of GTPs.

\section{Strip Condensation from Mirror Symmetry}

\label{section_mirror_symmetry}

We have argued that the combination of parallel legs that takes place in a  $(p,q)$ web to form a GTP naturally maps to the identification of the corresponding zig-zag paths. But it is important to understand whether this process goes beyond mere identification. Since these parallel zig-zags are the boundaries of an $\mathcal{N}=2$ fractional brane, we have argued that their identification physically corresponds to bringing them together, sending the gauge coupling of faces on the $\mathcal{N}=2$ strip to infinity. In this section we use mirror symmetry to show that this is indeed the case. 

Let us first quickly review some basics of the mirror geometry construction for toric CY$_3$'s \cite{Hori:2000kt,Feng:2005gw}. Given a toric CY$_3$ $\mathcal{M}$, its geometry is encoded in its toric diagram $\Delta$. We can associate to $\Delta$ a Newton polynomial by assigning to each point with coordinates $(m,n)$ the monomial $x^my^n$, with $x,y\in\mathbb{C}^*$. The resulting polynomial takes the form
\beq
    P(x,y)=\sum_{(m,n)\in \Delta\cap \mathbb{Z}^2} c_{m,n}x^my^n\,,
\label{general_form_Newton_polynomial}
\eeq
where the $c_{m,n}$ are complex coefficients. We can always freely fix 3 of them, by rescaling $x$ and $y$ and an overall $SL(2,\mathbb{Z})$ transformation.

The mirror geometry $\mathcal{W}$ is a double fibration over the complex plane $\mathbb{C}$ parametrized by $W$ as follows\footnote{Strictly speaking, this construction only works for $\Delta$'s with at least one internal point.}
\begin{equation}\label{eq:mirror_general}
\mathcal{W}=\begin{cases} P(x,y)=W\,,\\ uv=W\,. \end{cases}
\end{equation}

For $W\ne 0$, $uv=W$ parametrizes a copy of $\mathbb{C}$, with an $S^1$ associated to the phase of the free complex coordinate (which can be taken to be either $u$ or $v$). This $S^1$ collapses to zero size at $W=0$. In addition, For every $W$, $P(x,y)=W$ defines a Riemann surface, whose genus is the number of internal points of $\Delta$. Next, let us consider the critical points of $P$, namely the the points $(x_I,y_I)$ such that
\begin{equation}
    \frac{\partial P}{\partial x}\Big|_{(x_I,y_I)}= \frac{\partial P}{\partial y}\Big|_{(x_I,y_I)}=0\,.
\end{equation}
On top of every critical value $W_I=P(x_I,y_I)$, there is an $S^1$ in the Riemann surface that pinches off. Thus, the segment $[0,W_I]$ on the $W$-plane connecting the origin to $W_I$, combined with the $S^1\times S^1$ associated to the two cycles discussed above (one on the Riemann surface, which pinches off at $W=W_I$, and the other one on the $uv$-plane, which pinches off at $W=0$) defines a topological $S_I^3$. The number of critical points, and therefore of such $S^3_I$'s, is equal to the area of the toric diagram (measured in terms of minimal triangles). Gauge groups in the quiver correspond to D6-branes wrapping these $S^3$'s, with chiral fields arising at their intersections. The Riemann surface at the origin is particularly important, since the $S^3$'s intersect on top of it. We will refer to it as $\Sigma$.
\beq
\Sigma: \ \ \ \ P(x,y)=0 \, .
\eeq
Interestingly, $\Sigma$ is also the analogue of the Seiberg-Witten curve for the corresponding $5d$ theory compactified on a circle \cite{Aharony:1997bh}.

Two natural projections of $\Sigma$ are usually considered.

\paragraph{Amoeba.} 
The {\it amoeba} projection is defined as follows
\beq
(x,y) \to (\log |x|,\log|y|) \,.
\eeq
The amoeba can be viewed as a thickened version of the corresponding $(p,q)$ web. Equivalently, we can think about the $(p,q)$ web as the skeleton or tropical limit of the amoeba.

\paragraph{Coamoeba.} 

The coamoeba is the projection onto the angular parts of $(x,y)$, i.e.
\beq
(x,y) \to (\arg x,\arg y) \,.
\eeq
The coamoeba lives on $\mathbb{T} ^2$ and, for toric diagrams, its tropical limit is the corresponding brane tiling. Indeed, this approach provides a practical approach—albeit with some practical limitations, which we discuss below—for determining brane tilings from geometry for arbitrary coefficients in the Newton polynomial. In theories with multiple toric phases, varying the coefficients in the Newton polynomial captures their different brane tilings.

\subsection{Continuous Interpolation from Toric Diagrams to GTPs: The Fate of $\mathcal{N}=2$ Strips}

Turning a toric diagram into a GTP corresponds to the identification of parallel zig-zag paths in the corresponding brane tiling. To understand this operation more precisely, it is essential to clarify how the identification is implemented. For example, one could imagine the procedure could involve nontrivial “folding” of $\mathbb{T}^2$, potentially producing a graph embedded in a more intricate—and possibly singular—surface. However, we have proposed that the process amounts to continuously bringing the parallel zig-zag paths together, thereby collapsing the faces contained within the $\mathcal{N}=2$ strip between them. In this section, we show that mirror symmetry offers a natural framework to track this process in a controlled manner and confirm this picture.

The mirror geometry associated with GTPs was studied in \cite{Arias-Tamargo:2024fjt}, where it was shown to work as for ordinary toric diagrams, with the main difference being that the coefficients of $P(x,y)$ are subject to constraints. Heuristically, only the coefficients corresponding to black dots are independent.\footnote{Three of which can be fixed, as previously explained for toric diagrams.} The precise relations among the coefficients can be determined by generating the GTP through algebraic polytope mutation from a toric diagram \cite{GalkinUsnich,akhtar2012minkowski,Arias-Tamargo:2024fjt}.

To illustrate our ideas, let us consider the GTP obtained by polytope mutation of $dP_0$ shown in \fref{Toric_dP0_mutation_triangulated}. Using the algebraic version of polytope mutation, it was determined in \cite{Arias-Tamargo:2024fjt} that the Newton polynomial for this GTP has the general form
\beq
P(x,y)={1\over y}+1+\left({1\over x} + 2 c_1 + c_1^2 x \right) y^2
\label{Newton_polynomial_GTP_example}
\eeq
with $c_1 \in \mathbb{C}$ an arbitrary coefficient. In the general notation of \eqref{general_form_Newton_polynomial} for coefficients, $c_{0,2}$ and $c_{1,2}$ are related, while $c_{0,1}$ vanishes. As for the $dP_0$ toric diagram it is connected to by mutation, it has a single independent coefficient.

It is well known that there exists an algorithm that, in principle, allows the reconstruction of the the brane tiling from the coamoeba. Roughly speaking, the coamoeba projection induces a two-color decomposition of $\mathbb{T}^2$ \cite{Feng:2005gw}: one color corresponds to nodes and edges, while the other corresponds to faces (see also \cite{Seong:2024wkt} for a recent discussion).\footnote{This algorithm is closely related to the {\it Fast Inverse Algorithm}, who reconstruct brane tilings from paths with fixed winding numbers on $\mathbb{T}^2$ representing zig-zag paths. However, the procedure that is commonly understood as the Fast Inverse Algorithm, is combinatorial and does not explicitly use the Newton polynomial.} For suitable choices of the parameters in $P(x,y)$, the regions associated with nodes become particularly transparent, appearing as domains connected at vertices or through narrow paths that translate into the edges of the tiling. For other parameter choices, however, identifying the corresponding brane tiling can be considerably more subtle. 

To avoid this difficulty, we begin with a choice of parameters for which the polytope is viewed as an ordinary toric diagram, chosen such that the underlying brane tiling is clearly identifiable. We will then continuously track the evolution of the coamoeba as the parameters are varied towards the GTP limit given by \eqref{Newton_polynomial_GTP_example}. This approach allows us to follow the fate of the $\mathcal{N}=2$ strip as we approach the GTP. As discussed in Section \sref{subsection_parallel_zig_zags_fractional_branes}, any pair of parallel zig-zag paths defines two complementary regions, each corresponding to an $\mathcal{N}=2$ fractional brane. However, only the collapse of one of these regions corresponds to the GTP and, for example, yields the correct counting of gauge groups, as illustrated in the examples presented in the previous section. As we will see, mirror symmetry uniquely determines which of the two regions is the one that shrinks.\footnote{As mentioned earlier, there are non-generic cases in which an $\mathcal{N}=2$ strip and its complement contain the same number of gauge groups. In such cases, the resulting condensations cannot be distinguished solely by the number of surviving gauge groups. Mirror symmetry, however, also identifies the condensing strip in these cases.}

We will organize the initial and final parameters into vectors of the form $\vec{c}=(c_{0,-1},c_{0,0},c_{0,1},c_{0,2},c_{-1,2},c_{1,2})$ and consider the following initial and final values:
\beq
\begin{array}{ccl}
\vec{c}_0 & = & (2,i, 2 + 0.2 \, i, 2, 1 + i, 1 + i) \\[.15cm]
\vec{c}_f & = & (1, 1, 0, 2, 1, 1)
\end{array}
\label{initial_and_final_coefficients}
\eeq
where the final parameters in $\vec{c}_f$ are of the form \eqref{Newton_polynomial_GTP_example}, corresponding to the GTP we are interested in. We will consider a linear trajectory interpolating between $\vec{c}_0$ and $\vec{c}_f$, divided into five equal intervals. Namely, we will consider the following sequence of coefficients
\beq
\vec{c}(n)={(n-5)\over 5} \, \vec{c}_0 + {n \over 5} \, \vec{c}_f  \qquad , \qquad n=0,\ldots,5
\label{sequence_coefficients_interpolation}
\eeq

\begin{figure}[H]
\centering
\includegraphics[width=12cm]{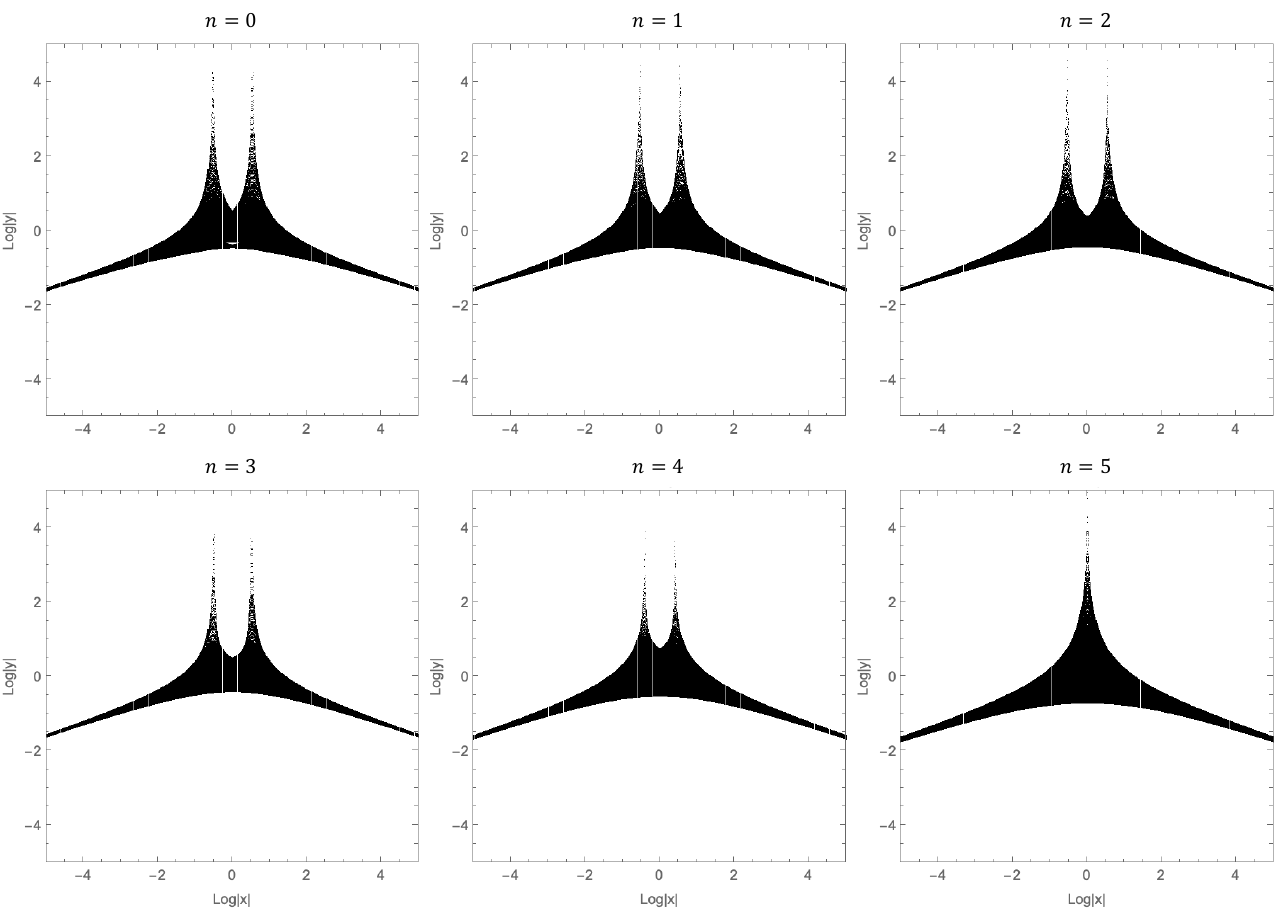}
\caption{Evolution of the amoeba as the coefficients in the Newton polynomial are varied from $\vec{c}_0$ to $\vec{c}_f$ in \eqref{initial_and_final_coefficients}, i.e. from those of a standard toric diagram to those satisfying the constraints of the GTP in \fref{Toric_dP0_mutation_triangulated}.}
\label{amoebae_interpolation}
\end{figure}

\fref{amoebae_interpolation} shows the resulting sequence of amoebae. As expected, the two parallel legs are initially separated and gradually come closer until merging into a single one for the GTP.\footnote{When regarded as a toric diagram, the GTP in \fref{Toric_dP0_mutation_triangulated} has two internal points. As a result, we generically expect two holes in the corresponding amoeba. However, depending on the choice of coefficients, those holes may not be manifest, as in \fref{amoebae_interpolation}.} In the first steps, the amoeba changes only slightly. This is just a result of the particular trajectory in the space of coefficients \eqref{sequence_coefficients_interpolation} that we picked in this particular example. We will observe a similar behavior for the coamoeba.

\begin{figure}[H]
\centering
\includegraphics[width=\textwidth]{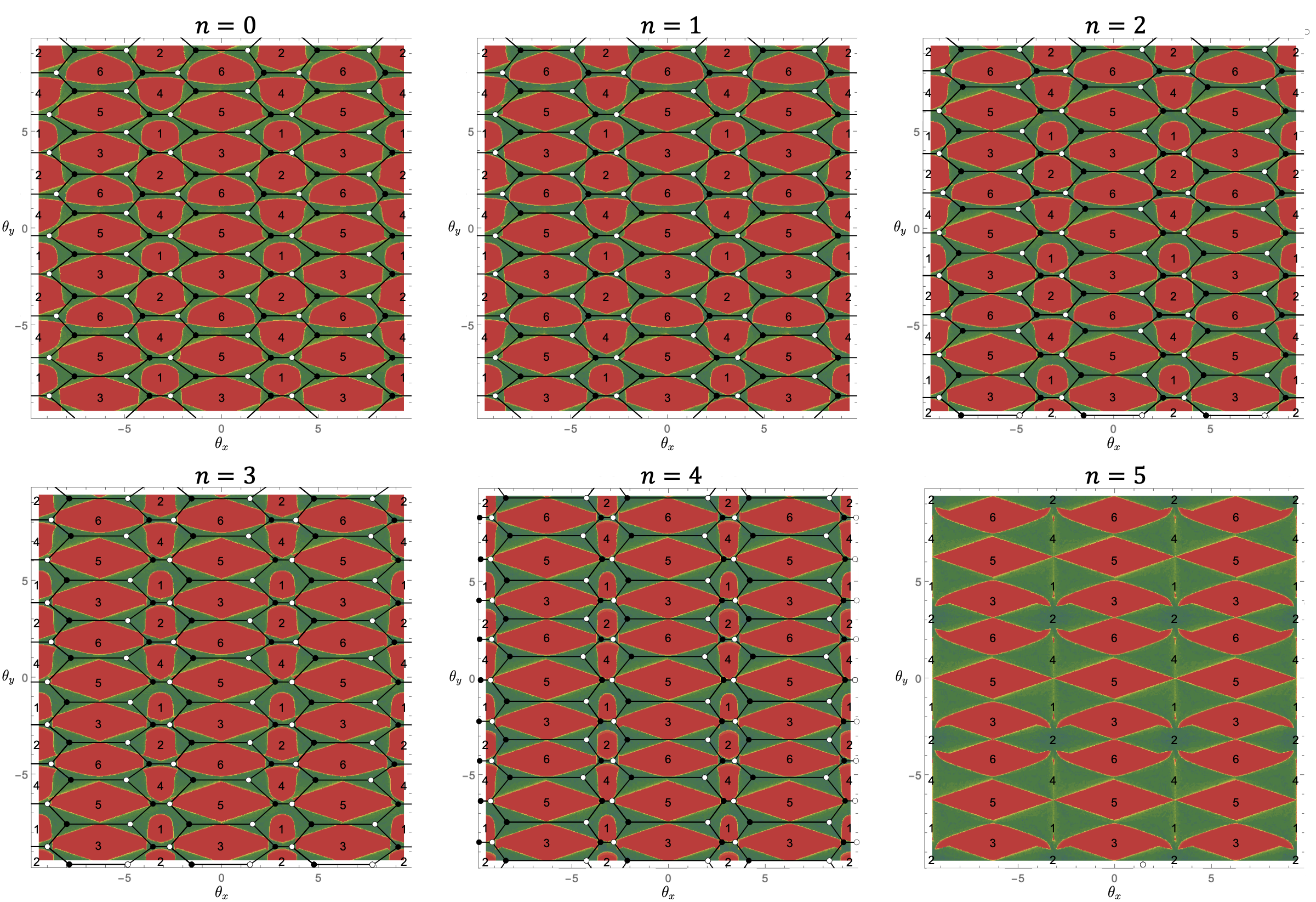}
\caption{Evolution of the coamoeba and the corresponding brane tiling as the coefficients in the Newton polynomial are varied from $\vec{c}_0$ to $\vec{c}_f$ in \eqref{initial_and_final_coefficients}. We observe how the $\mathcal{N}=2$ strip consisting of gauge groups 1, 2 and 4 continuously shrinks to zero size. The figure shows a region larger than a single unit cell.}
\label{Coamoebae}
\end{figure}

Let us now consider the corresponding coamoeabae, which are shown in \fref{Coamoebae}. At every step, we have superimposed the corresponding brane tiling. The brane tiling for the first step is easily identifiable and, as expected, it agrees with the one for the $\mathbb{C}^3/(\mathbb{Z}_2 \times \mathbb{Z}_3)$ orbifold we previously considered in \fref{tiling_Z2_Z3}. Since the coamoeba changes slightly at each step, we can continuously track the evolution of the brane tilings. Interestingly, we observe that the $\mathcal{N}=2$ strip consisting of nodes 1, 2, and 4 shrinks until it disappears at the GTP point, precisely as we proposed.\footnote{Of course, we have labeled the faces of the tiling to agree with \fref{tiling_Z2_Z3}.} Therefore, mirror symmetry confirms our interpretation of zig-zag path identification as the result of $\mathcal{N}=2$ strip condensation. We have not drawn a brane tiling for the last step in \fref{Coamoebae}, since the resulting theory is not described by one. In more general examples, where the strip and its complement are not as symmetric as in the present case, mirror symmetry likewise identifies which of the two is the one that shrinks. The same approach applies to examples with more than two identified zig-zags, i.e. with multiple consecutive white dots on a side of the GTP, and their corresponding strips.

\begin{figure}[H]
\centering
\includegraphics[width=10cm]{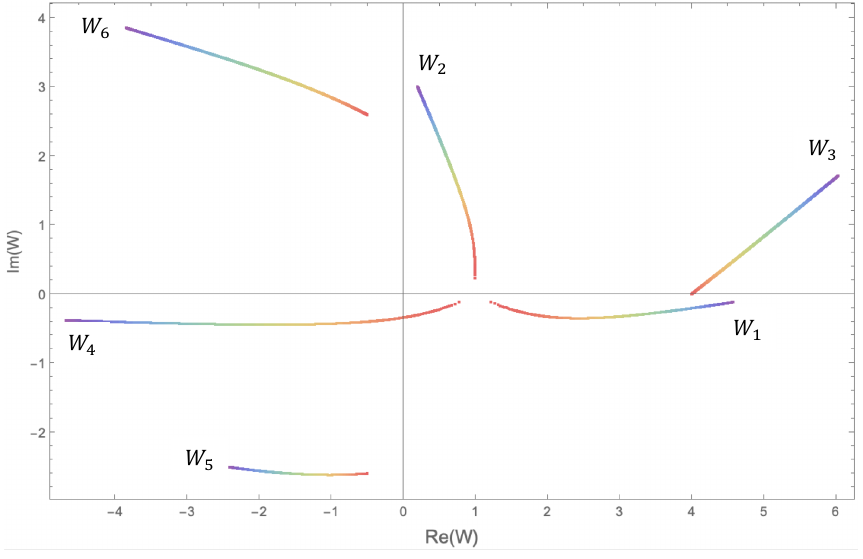}
\caption{Evolution of the critical values associated to each of the gauge groups as the coefficients evolve from a toric diagram (violet) to a GTP (red).}
\label{evolution_critical_points}
\end{figure}

The disappearance of gauge groups 1, 2 and 4 can also be observed in terms of the evolution of the critical values $W_I$ associated to the critical points $(x_I,y_I)$. This is shown in \fref{evolution_critical_points}, starting at violet from $\vec{c}_0$ and going towards red for $\vec{c}_f$. As in \eqref{sequence_coefficients_interpolation}, we have considered a linear interpolation connecting $\vec{c}_0$ and $\vec{c}_f$, but dividing it in many more intervals to show the continuous trajectories. The critical points for the shrinking faces, $W_1$, $W_2$ and $W_4$, approach $W=1$ and disappear. The other critical points, distribute symmetrically at distance 3 from $W=1$: namely $W_3 \to 4$, $W_6 \to 1+ 3 \, e^{i{2 \pi/3}}$ and $W_5 \to 1+ 3 \, e^{-i{2 \pi/3}}$.\footnote{Setting the constant term in \eqref{Newton_polynomial_GTP_example} to 0 instead of 1 shifts the trajectories such that $W_1$, $W_2$, and $W_4$ approach the origin before disappearing, while $W_3$, $W_6$, and $W_5$ converge to a configuration with $\mathbb{Z}_3$ symmetry around the origin: $W_3 \to 3$, $W_6 \to 3 \, e^{i{2 \pi/3}}$ and $W_5 \to 3 \, e^{-i{2 \pi/3}}$.}

\section{Quivers for GTPs: Strip Condensation and Confinement}

\label{section_confinement}

The identification of parallel zig-zags corresponds to shrinking the $\mathcal{N}=2$ strips between them to zero size. Physically, this implies sending the gauge coupling of faces within the strip to infinity and their confinement. In this section, we provide further evidence for this proposal. 

When a gauge group confines, the elementary chiral fields charged under it are no longer the appropriate degrees of freedom, and one should instead describe the low energy theory in terms of gauge invariant composite operators. We will restrict our attention to mesons of length 2, meaning mesons consisting of precisely two chiral fields connected through a single confining node. In principle, it is possible for confinement to generate longer mesons, namely involving more than two chiral fields charged under multiple confining gauge groups. While we currently lack a first-principles argument to rule out the necessity of longer mesons, we will see that length-2 mesons alone already lead to a compelling picture. We remain agnostic about the precise form of the superpotential of the confined theory. In particular, since the global symmetry of the theory changes, there is no reason to expect that the superpotential is obtained by a naive substitution of elementary fields by their meson counterparts in the original superpotential. For this reason, we will not make any assumptions about whether particular fields acquire masses or not.

With these ingredients, we revisit most of the examples analyzed in Section \sref{section_parallel_zigzas_N2_branes}. Remarkably, we find that restricting to length-2 mesons is sufficient to produce a quiver that contains the quiver before the mutation, differing from it at most by vector pairs or adjoint fields.\footnote{We use the term ``at most" to indicate that, given that we do not know the superpotential, we cannot determine whether such fields are massive or not.} Therefore, the structure of the resulting theories is consistent with being connected to the original ones by relevant superpotential deformations (either mass or non-mass), which is the case for HW transitions/polytope mutations connecting toric diagrams \cite{Bianchi:2014qma,Cremonesi:2023psg}. This provides non-trivial evidence that the confinement of the $\mathcal{N}=2$ strip correctly captures the physics of the GTP.\footnote{Strictly speaking, in this section we will show that length-2 mesons are sufficient in examples involving $T_2$-cones. It is possible that theories with $T_n$-cones for $n>2$ require the inclusion of longer mesons. We plan to revisit this question in future work.}

\subsection*{Example 1: Mutation of $dP_0$}

\begin{figure}[ht!]
\centering
\includegraphics[width=4cm]{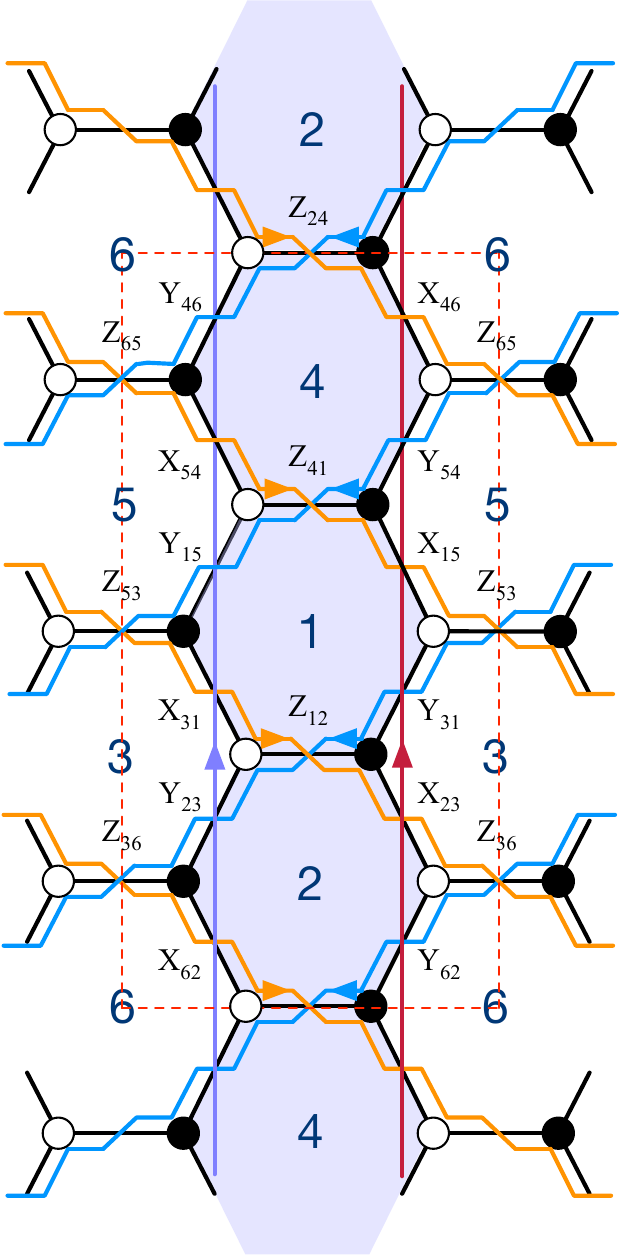}
\caption{Brane tiling from \fref{tiling_Z2_Z3}, with edge labels indicating the corresponding fields to facilitate the analysis of confinement.}
\label{Tiling_Z2_Z3_with_labels}
\end{figure}

For quick reference, in \fref{Tiling_Z2_Z3_with_labels} we show again the brane tiling when interpreting the GTP as a toric diagram. For every edge, we have indicated the corresponding field. 

Let us consider the fields after confinement.\footnote{Given our assumptions, the fields that live on the strip and are charged under pairs of confining gauge groups, namely $Z_{24}$, $Z_{41}$ and $Z_{12}$, will not play any role in our analysis.} We restrict to length-2 mesons. We use different colors to distinguish mesons arising from fields charged under different confining gauge groups, making them easier to identify in the final quiver, which we show in \fref{quivers_condensation_dP0}.
\begin{itemize}
\item Fields that are not on the strip, namely fields neutral under the confined gauge groups (black):
\beq
Z_{65} \ \, \ \ Z_{53} \ \, \ \ Z_{36}  
\eeq

\item Mesons with fields charged under face 4 \textcolor{red}{(red)}:
\beq
\begin{array}{ccc}
M_{56}^{(1)}=X_{54} X_{46} & \ \ \ \ & M_{56}^{(2)}=X_{54} Y_{46} \\[.15cm]
M_{56}^{(3)}=Y_{54} X_{46} & \ \ \ \ & M_{56}^{(4)}=Y_{54} Y_{46}
\end{array}
\eeq

\item Mesons with fields charged under face 1 \textcolor{mygreen}{(green)}:
\beq
\begin{array}{ccc}
M_{35}^{(1)}=X_{31} X_{15} & \ \ \ \ & M_{35}^{(2)}=X_{31} Y_{15} \\[.15cm]
M_{35}^{(3)}=Y_{31} X_{15} & \ \ \ \ & M_{35}^{(4)}=Y_{31} Y_{15} 
\end{array}
\eeq

\item Mesons with fields charged under face 2 \textcolor{blue}{(blue)}:
\beq
\begin{array}{ccc}
M_{63}^{(1)}=X_{62} X_{23} & \ \ \ \ & M_{63}^{(2)}=X_{62} Y_{23} \\[.15cm]
M_{63}^{(3)}=Y_{62} X_{23} & \ \ \ \ & M_{63}^{(4)}=Y_{62} Y_{23}
\end{array}
\eeq
\end{itemize}

\fref{quivers_condensation_dP0}.a shows the quiver for the toric diagram before mutation, namely the quiver for $dP_0$ (see e.g. \cite{Feng:2002zw}). \fref{quivers_condensation_dP0}.b shows the quiver for the surviving nodes and containing the fields discussed above. This is the quiver obtained by confinement in the theory of the mutated GTP interpreted as a toric diagram. Interestingly, the two quivers differ at most by vector pairs. This is consistent with the mutated theory being connected to the original one by a relevant deformation.

\begin{figure}[H]
\centering
\includegraphics[width=8cm]{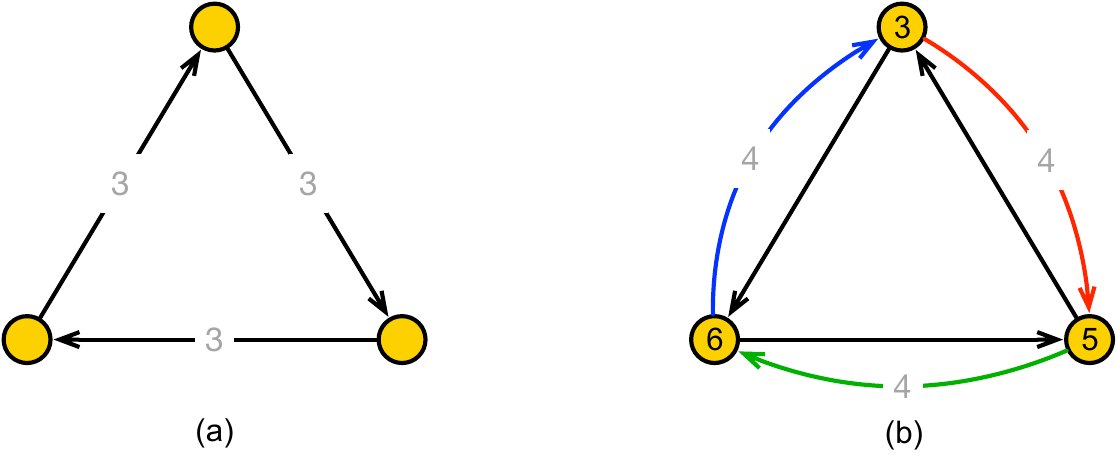}
\caption{Quivers for the polytope mutation of $dP_0$ shown in \fref{Toric_dP0_mutation_triangulated}. (a) Initial quiver for $dP_0$. (b) Quiver associated with the GTP, obtained by confining an $\mathcal{N}=2$ strip.}
\label{quivers_condensation_dP0}
\end{figure}

\subsubsection*{Example 2: Mutation of $dP_1$}

\begin{figure}[H]
\centering
\includegraphics[width=7cm]{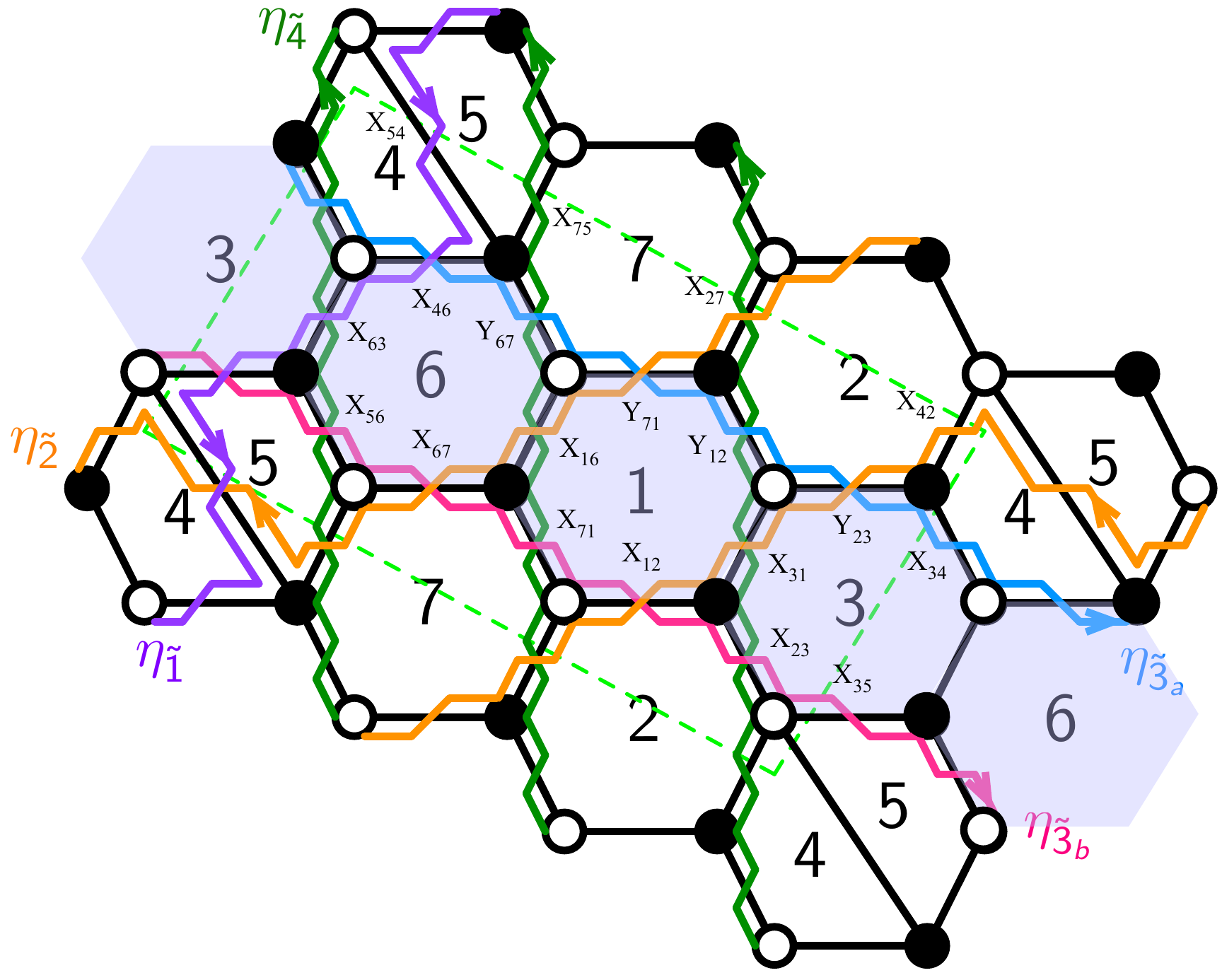}
\caption{Brane tiling from \fref{Tiling_mutation_dP1}, with edge labels indicating the corresponding fields.}
\label{Tiling_mutation_dP1_with_labels}
\end{figure}

\fref{Tiling_mutation_dP1_with_labels} shows again the brane tiling when interpreting the GTP as a toric diagram with its edges labeled.

Proceeding as before, the fields after confinement are:

\begin{itemize}
\item Fields not on the strip (black):
\beq
X_{42} \ \, \ \ X_{27} \ \, \ \ X_{75} \ \, \ \ X_{54} 
\eeq

\item Mesons with fields charged under face 6 \textcolor{red}{(red)}:
\beq
\begin{array}{ccc}
M_{57}^{(1)}=X_{56} X_{67} & \ \ \ \ & M_{57}^{(2)}=X_{56} Y_{67} \\[.15cm]
M_{47}^{(1)}=X_{46} X_{67} & \ \ \ \ & M_{47}^{(2)}=X_{46} Y_{67} 
\end{array}
\eeq

\item Mesons with fields charged under face 1 \textcolor{mygreen}{(green)}:
\beq
\begin{array}{ccc}
M_{72}^{(1)}=X_{71} X_{12} & \ \ \ \ & M_{72}^{(2)}=X_{71} Y_{12} \\[.15cm]
M_{72}^{(3)}=Y_{71} X_{12} & \ \ \ \ & M_{72}^{(4)}=Y_{71} Y_{12} 
\end{array}
\eeq

\item Mesons with fields charged under face 3 \textcolor{blue}{(blue)}:
\beq
\begin{array}{ccc}
M_{24}^{(1)}=X_{23} X_{34} & \ \ \ \ & M_{24}^{(2)}=Y_{23} X_{34} \\[.15cm]
M_{25}^{(1)}=X_{23} X_{35} & \ \ \ \ & M_{25}^{(2)}=Y_{23} X_{35} 
\end{array}
\eeq
\end{itemize}

\begin{figure}[H]
\centering
\includegraphics[width=9cm]{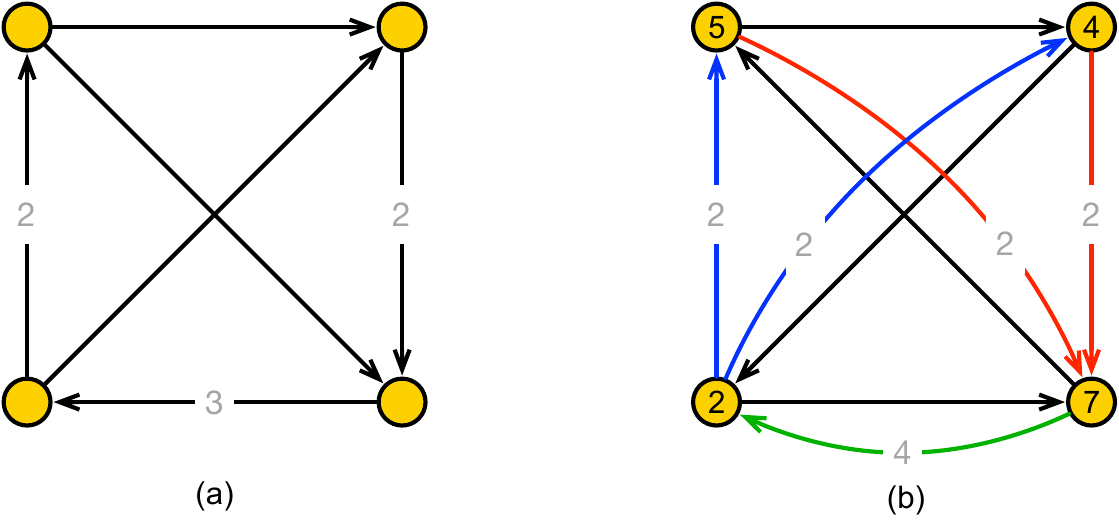}
\caption{Quivers for the polytope mutation of $dP_1$ shown in \fref{Toric_dP1}. (a) Initial quiver for $dP_1$. (b) Quiver associated with the GTP, obtained by confining an $\mathcal{N}=2$ strip.}
\label{quiver_dP1_and_mutation}
\end{figure}

\fref{quiver_dP1_and_mutation}.a shows the quiver for $dP_1$, the theory before mutation \cite{Feng:2002zw}. \fref{quiver_dP1_and_mutation}.b shows the quiver obtained by confinement from the mutated GTP interpreted as a toric diagram. Once again, the two quivers differ at most by vector pairs.

\subsubsection*{Examples 3 and 4: Mutation of $dP_2$}

The following examples are particularly interesting because, as explained in \cite{Franco:2023flw}, they are generated by starting from the two toric phases of $dP_2$. Each of these models is connected to a specific toric phase of $dP_2$ through the algorithm of \cite{Higashitani:2019vzu}. These examples are also interesting because the intricate structure of the tilings outside the strips.

\begin{figure}[H]
\centering
\includegraphics[width=4.7cm]{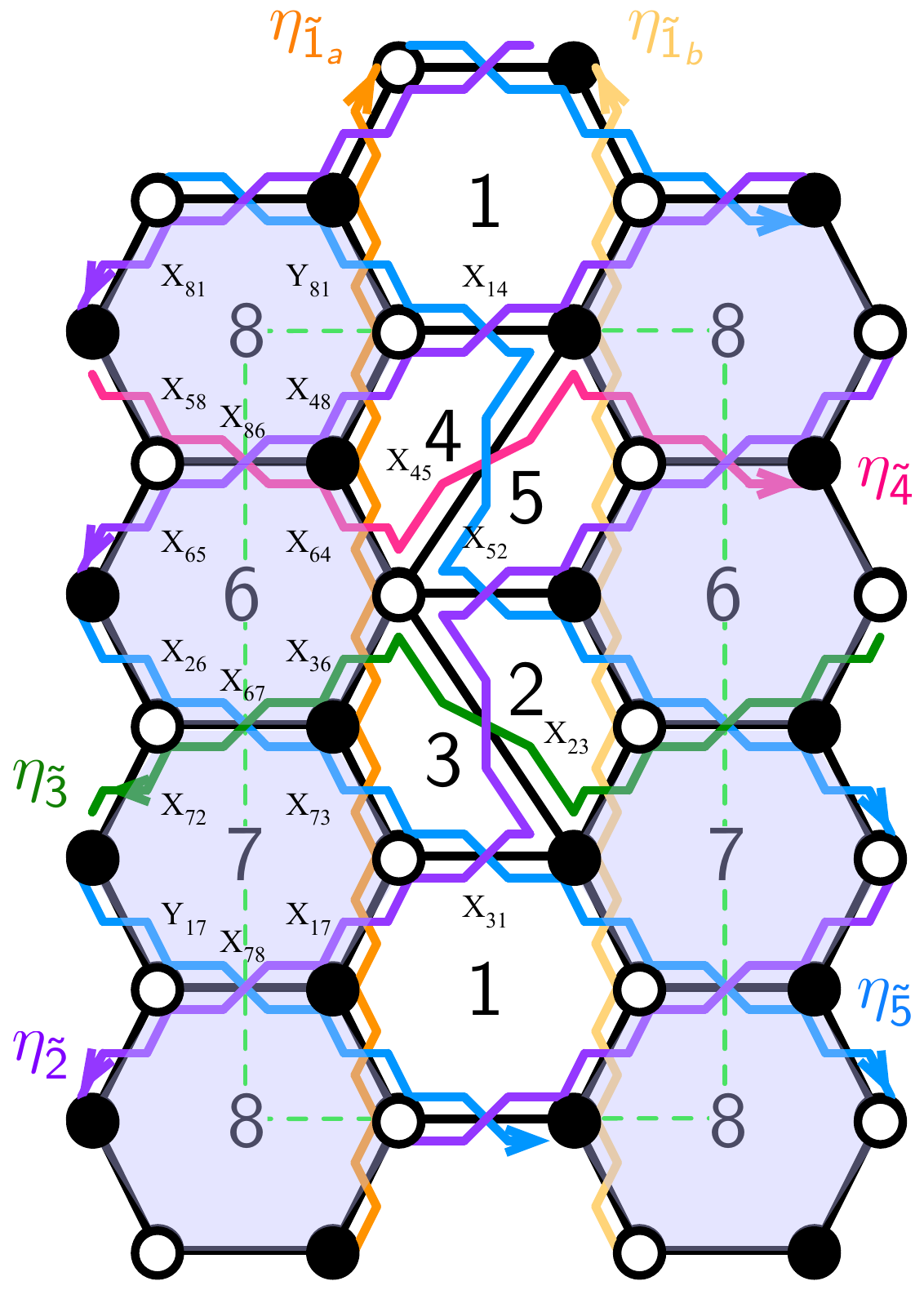}
\caption{Brane tiling from \fref{Tiling_mutation_dP2_1}, with edge labels indicating the corresponding fields.}
\label{Tiling_mutation_dP2_1_labeled}
\end{figure}

\fref{Tiling_mutation_dP2_1_labeled} shows the brane tiling for the first model \cite{Franco:2023flw}.

The fields after confinement are:

\begin{itemize}
\item Fields not on the strip (black):
\beq
X_{14} \ \, \ \ X_{45} \ \, \ \ X_{52} \ \, \ \ X_{23} \ \, \ \ X_{31} 
\eeq

\item Mesons with fields charged under face 8 \textcolor{red}{(red)}:
\beq
\begin{array}{ccc}
M_{41}^{(1)}=X_{48} X_{81} & \ \ \ \ & M_{41}^{(2)}=X_{48} Y_{81} \\[.15cm]
M_{51}^{(1)}=X_{58} X_{81} & \ \ \ \ & M_{51}^{(2)}=X_{58} Y_{81}
\end{array}
\eeq

\item Mesons with fields charged under face 6 \textcolor{mygreen}{(green)}.
\beq
\begin{array}{ccc}
M_{24}=Y_{26} X_{64} & \ \ \ \ & M_{25}=X_{26} X_{65} \\[.15cm]
M_{34}=X_{36} X_{64} & \ \ \ \ & M_{35}=X_{36} X_{65} 
\end{array}
\eeq

\item Mesons with fields charged under face 7 \textcolor{blue}{(blue)}:
\beq
\begin{array}{ccc}
M_{12}^{(1)}=X_{17} X_{72} & \ \ \ \ & M_{12}^{(2)}=Y_{17} X_{72} \\[.15cm]
M_{13}^{(1)}=X_{17} X_{73} & \ \ \ \ & M_{13}^{(2)}=Y_{17} X_{73}
\end{array}
\eeq
\end{itemize}

\begin{figure}[H]
\centering
\includegraphics[width=10cm]{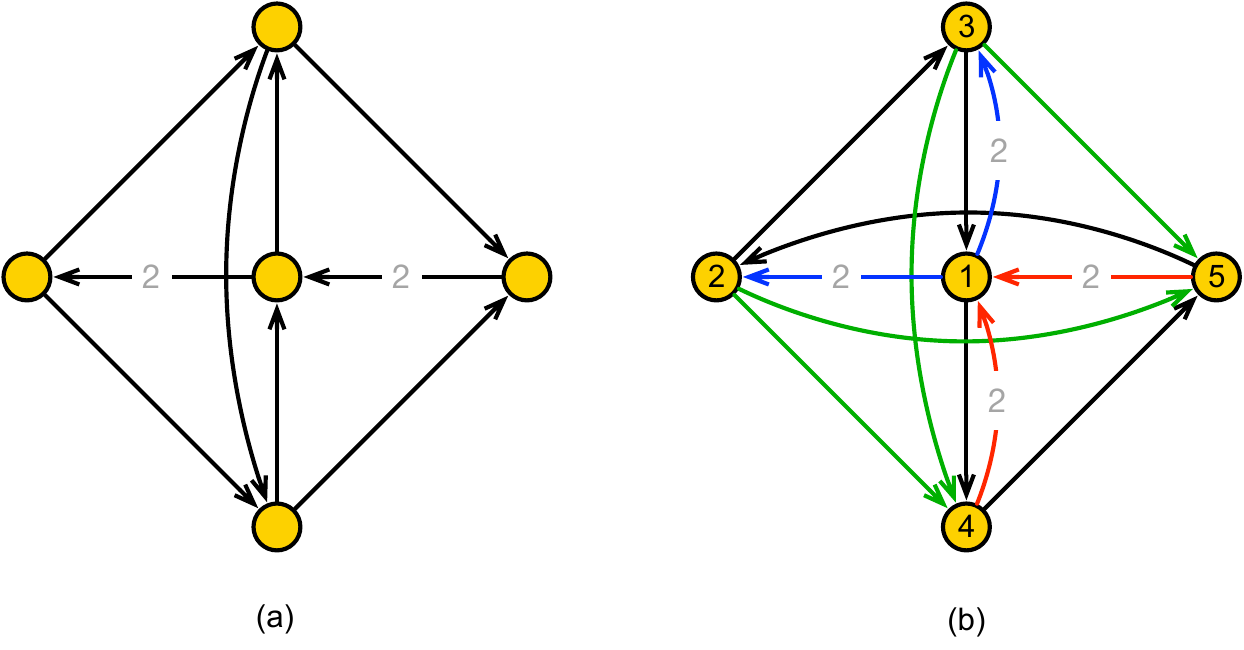} 
\caption{Quivers for the polytope mutation of $dP_2$ shown in \fref{Toric_dP2}. (a) Initial quiver for Model 2 of $dP_2$. (b) Quiver associated with the GTP, obtained by confining an $\mathcal{N}=2$ strip in the brane tiling constructed using the algorithm in \cite{Higashitani:2019vzu}.}
\label{quivers_condensation_dP0}
\label{quiver_dP2_1_and_mutation}
\end{figure}

\fref{quiver_dP2_1_and_mutation}.a shows the theory before mutation, which is often called Model 2 of $dP_2$ \cite{Feng:2002zw,Franco:2005rj}. \fref{quiver_dP2_1_and_mutation}.b shows the quiver obtained by confinement from the mutated GTP interpreted as a toric diagram. As in previous examples, the two quivers differ at most by vector pairs.

\begin{figure}[H]
\centering
\includegraphics[width=6.5cm]{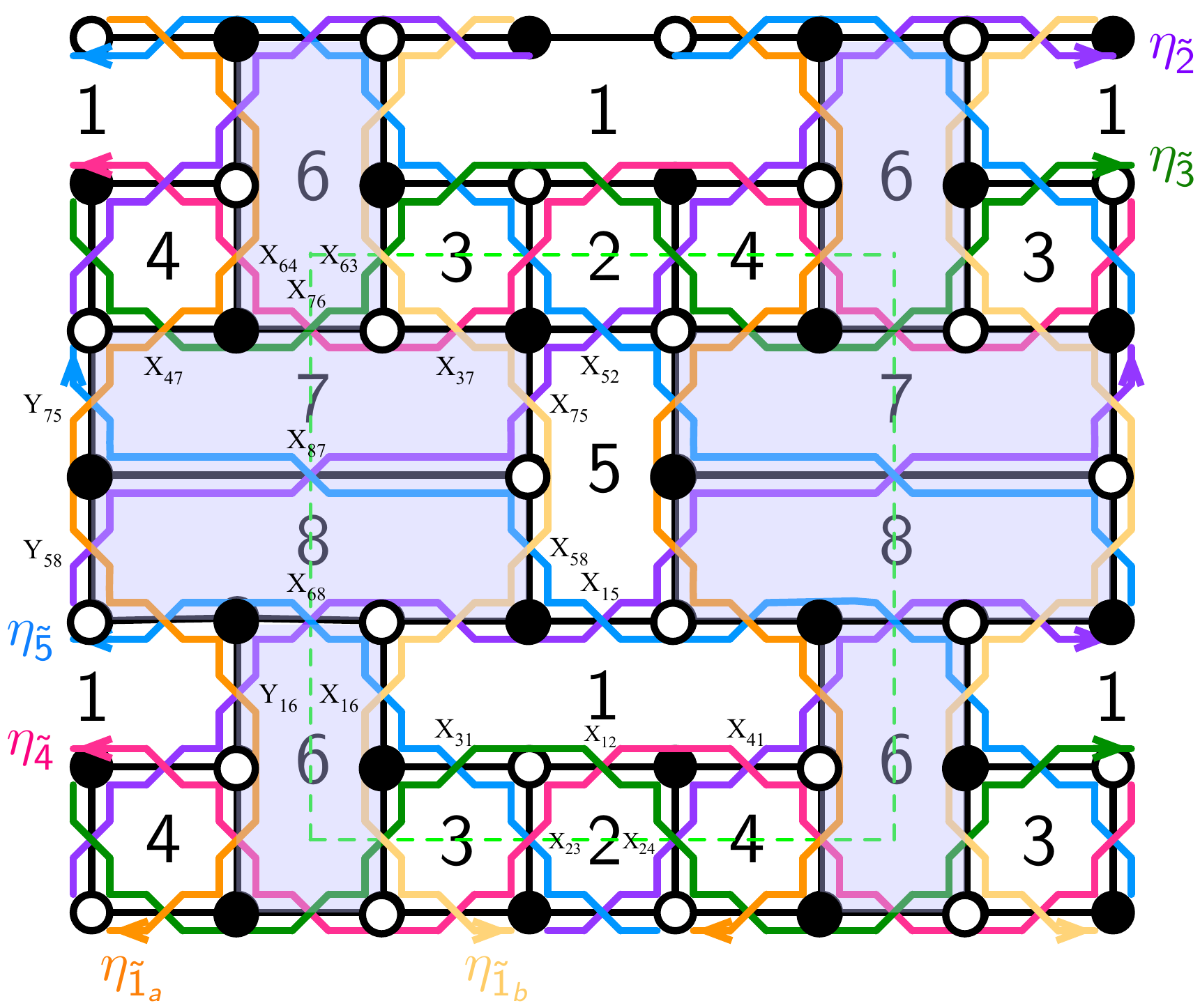}
\caption{Brane tiling from \fref{Tiling_mutation_dP2_2}, with edge labels indicating the corresponding fields.}
\label{Tiling_mutation_dP2_2_labeled}
\end{figure}

\fref{Tiling_mutation_dP2_2_labeled} shows the brane tiling for the second model \cite{Franco:2023flw}.

The fields after confinement are:

\begin{itemize}
\item Fields not on the strip (black):
\beq
X_{12} \ \, \ \ X_{24} \ \, \ \ X_{41} \ \, \ \ X_{15} \ \, \ \ X_{52} \ \, \ \ X_{23} \ \, \ \ X_{31} 
\eeq

\item Mesons with fields charged under face 7 \textcolor{red}{(red)}:
\beq
\begin{array}{ccc}
M_{35}^{(1)}=X_{37} X_{75} & \ \ \ \ & M_{35}^{(2)}=X_{37} Y_{75} \\[.15cm]
M_{45}^{(1)}=X_{47} X_{75} & \ \ \ \ & M_{45}^{(2)}=X_{47} Y_{75}
\end{array}
\eeq

\item Mesons with fields charged under face 8 \textcolor{mygreen}{(green)}:
\beq
\begin{array}{ccc}
M_{51}^{(1)}=X_{58} X_{81} & \ \ \ \ & M_{51}^{(2)}=X_{58} Y_{81} \\[.15cm]
M_{51}^{(3)}=Y_{58} X_{81} & \ \ \ \ & M_{51}^{(4)}=Y_{58} Y_{81}
\end{array}
\eeq

\item Mesons with fields charged under face 6 \textcolor{blue}{(blue)}:
\beq
\begin{array}{ccc}
M_{13}^{(1)}=X_{16} X_{63} & \ \ \ \ & M_{13}^{(2)}=Y_{16} X_{63} \\[.15cm]
M_{14}^{(1)}=X_{16} X_{64} & \ \ \ \ & M_{14}^{(2)}=Y_{16} X_{63}
\end{array}
\eeq
\end{itemize}

\begin{figure}[H]
\centering
\includegraphics[width=10cm]{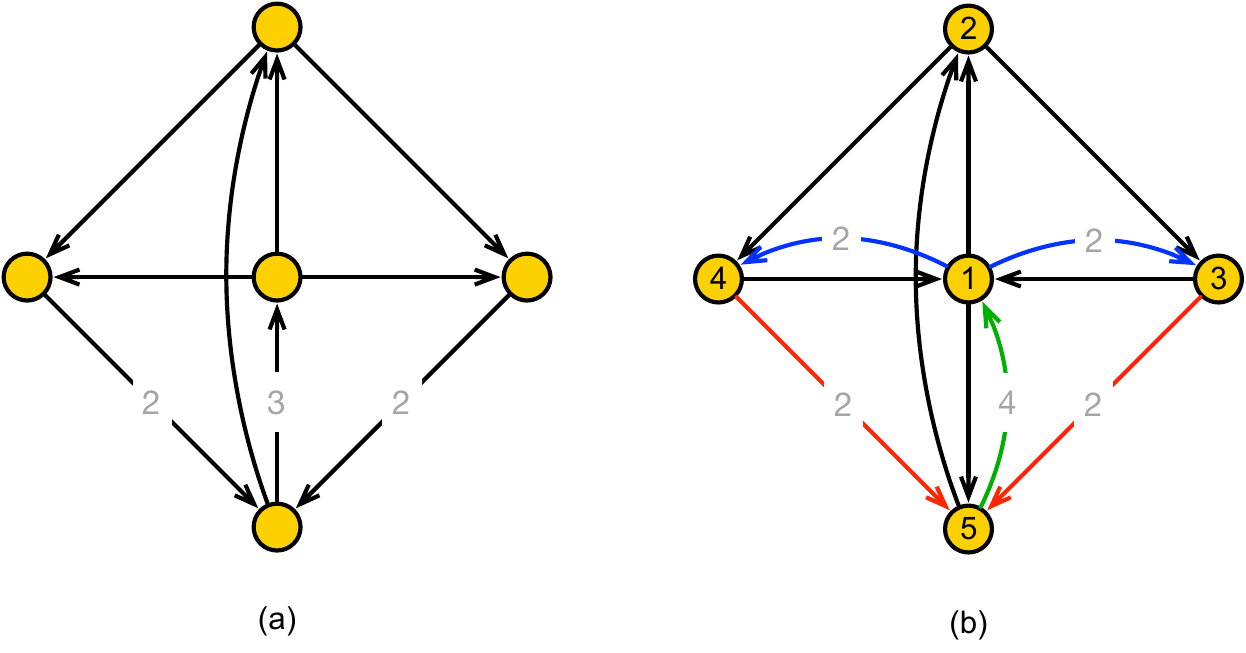} 
\caption{Quivers for the polytope mutation of $dP_2$ shown in \fref{Toric_dP2}. (a) Initial quiver for Model 1 of $dP_2$. (b) Quiver associated with the GTP, obtained by confining an $\mathcal{N}=2$ strip in the brane tiling constructed using the algorithm in \cite{Higashitani:2019vzu}.}
\label{quiver_dP2_2_and_mutation}
\end{figure}

\fref{quiver_dP2_2_and_mutation}.a shows the theory before mutation, often called Model 1 of $dP_2$ \cite{Feng:2002zw,Franco:2005rj}. \fref{quiver_dP2_2_and_mutation}.b shows the quiver obtained by confinement from the mutated GTP interpreted as a toric diagram. Once again, the two quivers differ at most by vector pairs, in agreement with our expectations.

\section{Conclusions and Future Directions}

\label{section_conclusions}

We have taken significant steps towards finding the generalization of brane tilings or, equivalently, the quivers with superpotentials corresponding to GTPs. We focused on GTPs connected to ordinary toric diagrams by polytope mutation or, equivalently, by a HW transition of the brane web.

Parallel 5-brane legs of a $(p,q)$ web define $\mathcal{N}=2$ fractional branes bounded by the associated parallel zig-zag paths. We proposed that terminating multiple 5-branes on a common 7-brane, which is the hallmark of GTPs, translates into bringing the corresponding zig-zag paths together, shrinking the $\mathcal{N}=2$ fractional branes to zero size in the process. We referred to this phenomenon as $\mathcal{N}=2$ strip condensation. By following the evolution for the Newton polynomial as it approaches the GTP point from a generic choice of coefficients, we showed that mirror symmetry indeed predicts strip condensation.

$\mathcal{N}=2$ strip condensation passes a network of additional non-trivial consistency checks that, taken together, paint a compelling picture. The expected number of gauge groups in the generalized brane tiling is that of the theory associated with a standard toric diagram connected to the GTP by polytope mutation. Alternatively, the number of gauge groups is given by the number of $T$-cones in a tessellation of the GTP. We showed that $\mathcal{N}=2$ produces the correct number of gauge groups for all examples previously considered in the literature as well as in a new infinite class of GTPs containing arbitrarily large $T$-cones. 

In addition, the shrinking of $\mathcal{N}=2$ fractional branes results in the corresponding gauge groups going to infinite coupling. We investigated the effect of confining these gauge groups and found that the resulting quivers coincide with those associated with the standard toric diagrams related to the GTPs by polytope mutation, up to, at most, the presence of extra chiral fields forming vector pairs or in adjoint represenations. This is remarkable, since it means that, in principle, the quiver theory for the GTP can be related to that of the toric diagram by a relevant deformation. This is known to be the case for polytope mutations connecting pairs of ordinary toric diagrams \cite{Bianchi:2014qma,Cremonesi:2023psg} and our results suggest that this connection extends to GTPs. Interestingly, we also studied examples with multiple toric phases and showed that this phenomenon continues to hold in these cases.\footnote{As mentioned earlier, it would be interesting to determine whether the length of the mesons to be considered depends on the size of the corresponding $T$-cones.}

At this point, determining the superpotential for these theories remains the most important open question. Once this is achieved, there are multiple directions for future work. Let us briefly mention some of them.

First, it would be interesting to determine whether there is a new class of objects with interesting combinatorial properties, what we would truly call generalized brane tilings, that capture the structure of these theories and their connection to the underlying geometry. More ambitiously, one could expect such objects to be connected by a generalization of untwisting to the yet to be determined extension of brane tilings describing the non-toric phases of the corresponding twin quivers.

It would also be interesting to study these theories as a new class of $4d$ $\mathcal{N}=1$ SCFTs on D3-branes probing the CY$_3$ defined by GTPs. In this context, it would be interesting to investigate whether a detailed correspondence can be established between quantities computed in the quantum field theory and in the geometry, such as superconformal $R$-charges and central charges, as has been done so successfully for toric theories \cite{Herzog:2002ih,Benvenuti:2004dy,Franco:2005sm}.

An intriguing connection with our recent work on the mutation-invariant Hilbert series \cite{Franco:2026bie} deserves further investigation. In that work, it was shown that, after choosing an origin in the toric diagram, a simple assignment of scaling dimensions to the fields of the underlying toric gauge theory reproduces the Ehrhart series associated with the corresponding GTP. From the perspective developed in this paper, this result suggests that ordinary brane tilings retain considerably more information about GTPs than previously appreciated. In particular, the picture proposed here naturally raises the possibility that the Ehrhart scaling is not an independent prescription, but rather the algebraic manifestation of the strip-condensation mechanism introduced in this work. If so, the generalized brane tilings proposed here and the mutation-invariant Hilbert series would provide complementary geometric and algebraic descriptions of the same underlying structure. Understanding this connection would provide a conceptual explanation for the origin of the Ehrhart scaling and further clarify how GTP data are encoded in ordinary brane tilings.

Cluster integrable systems for GTPs were recently studied in \cite{Kho:2026zwc}, where they were obtained by following the birational transformations that realize polytope mutations. For example, in these systems, the conserved quantities (Hamiltonians and Casimirs) inherit constraints from the restrictions on the coefficients of the Newton polynomial discussed in Section \sref{section_mirror_symmetry} (see \cite{Arias-Tamargo:2024fjt}). While this construction is extremely interesting, it would be desirable to derive such integrable systems directly from some quiver theory or generalized brane tiling via a generalization of Goncharov-Kenyon prescription. Various features of the quivers derived in this work indicate they are the right objects for this construction. First, the number of gauge groups remains the same as in the theory before polytope mutation. This counting produces the number of free variables that is naively expected from the number of independent conserved quantities in these integrable systems after taking into account constraints \cite{Kho:2026zwc} and assuming that these theories are described by objects living on $\mathbb{T}^2$. Moreover, as we noted in Section \sref{section_confinement}, the quivers differ from the ones before mutation by, at most, chiral fields in vector pairs and adjoint representations. Quite suggestively, the presence of this ``unoriented” extra matter would not affect the Poisson brackets of the integrable system. 
We plan to revisit some of these questions in future work.

\acknowledgments

We would like to thank the Simons Center for Geometry and Physics for its kind hospitality during the completion of this work. S.F. is supported by the U.S. National Science Foundation grant PHY-2412479. D.R.-G is supported in part by the Spanish national grant MCIU-22-PID2021- 123021NB-I00.

\bibliographystyle{JHEP}
\bibliography{mybib}

\end{document}